\documentclass[sigconf,nonacm]{acmart}

\usepackage{subfigure}
\usepackage{siunitx}
\usepackage{enumitem}

\usepackage{amsmath}

\usepackage{algorithm}
\usepackage[noend]{algpseudocode}
\let\ReturnInline\Return
\renewcommand{\Return}{\State\ReturnInline}
\algrenewcommand\algorithmicrequire{$\rhd$}
\algrenewcommand\algorithmicensure{$\square$}

\AtBeginDocument{%
  \providecommand\BibTeX{{%
    \normalfont B\kern-0.5em{\scshape i\kern-0.25em b}\kern-0.8em\TeX}}}

\setcopyright{acmcopyright}
\copyrightyear{2018}
\acmYear{2018}
\acmDOI{XXXXXXX.XXXXXXX}

\acmConference[Conference acronym 'XX]{Make sure to enter the correct
  conference title from your rights confirmation emai}{June 03--05,
  2018}{Woodstock, NY}

\newcommand{\ignore}[1]{}

\begin{document}

\title[DF* PageRank: Improved Incrementally Expanding Approaches for Updating PageRank on Dynamic Graphs]{DF* PageRank: Improved Incrementally Expanding Approaches for Updating PageRank on Dynamic Graphs}


\author{Subhajit Sahu}
\email{subhajit.sahu@research.iiit.ac.in}
\affiliation{%
  \institution{IIIT Hyderabad}
  \streetaddress{Professor CR Rao Rd, Gachibowli}
  \city{Hyderabad}
  \state{Telangana}
  \country{India}
  \postcode{500032}
}


\settopmatter{printfolios=true}

\begin{abstract}
PageRank is a widely used centrality measure that assesses the significance of vertices in a graph by considering their connections and the importance of those connections. Efficiently updating PageRank on dynamic graphs is essential for various applications due to the increasing scale of datasets. This technical report introduces our improved Dynamic Frontier (DF) and Dynamic Frontier with Pruning (DF-P) approaches. Given a batch update comprising edge insertions and deletions, these approaches iteratively identify vertices likely to change their ranks with minimal overhead. On a server featuring a 64-core AMD EPYC-7742 processor, our approaches outperform Static and Dynamic Traversal PageRank by $5.2\times$/$15.2\times$ and $1.3\times$/$3.5\times$ respectively - on real-world dynamic graphs, and by $7.2\times$/$9.6\times$ and $4.0\times$/$5.6\times$ on large static graphs with random batch updates. Furthermore, our approaches improve performance at a rate of $1.8\times$/$1.7\times$ for every doubling of threads.
\end{abstract}

\keywords{Parallel PageRank algorithm, Improved Dynamic Frontier approach}


\maketitle

\section{Introduction}
\label{sec:introduction}
Centrality metrics quantify the importance of nodes within a network based on link structures. PageRank \cite{rank-page99}, originally devised to rank web pages in search results, is one the most popular centrality metrics. It is based on the principle that pages receiving a greater number of high-quality links are of higher quality and, consequently, should be assigned higher ranks. Given the importance of such a metric, PageRank finds applications beyond web page ranking, including urban planning \cite{urban-zhang18}, traffic flow prediction \cite{traffic-kim15}, protein target identification \cite{banky2013equal}, evaluating the importance of brain regions \cite{zuo2012network}, identifying species crucial to environ\textit{mental} health \cite{allesina2009googling}, characterizing the properties of a software system \cite{chepelianskii2010towards}, and quantifying the scientific impact of researchers \cite{rank-senanayake15}. The growing availability of extensive interconnected / graph-based data has fueled substantial interest in parallel algorithms for computing PageRank \cite{rank-garg16, rank-nvgraph, rank-giri20, rank-guoqiang20, rank-li21, rank-sadi18, rank-sarma13}.\ignore{--- it has been implemented on multicore CPUs \cite{rank-garg16}, GPUs \cite{rank-nvgraph}, FPGAs \cite{rank-guoqiang20}, SpMV ASICs \cite{rank-sadi18}, CPU-GPU hybrids \cite{rank-giri20}, CPU-FPGA hybrids \cite{rank-li21}, and distributed systems \cite{rank-sarma13}.}

However, the dynamic nature of most real-world graphs, characterized by frequent edge insertions and deletions, poses challenges for recomputing PageRank from scratch, especially when dealing with small, rapid changes \cite{agarwal2012real, barros2021survey}. To address this, existing strategies instead iterate from ranks of vertices obtained in a previous snapshot of the graph, thereby reducing the required number of iterations for convergence. To further minimize the runtime needed, it is necessary to recompute only the ranks of vertices that are likely to change. One prevalent approach involves identifying reachable vertices from the updated regions of the graph and limiting processing to these vertices \cite{rank-desikan05, kim2015incremental, rank-giri20, sahu2022dynamic}. However, marking all reachable vertices as affected, even for minor rank changes, is likely to result in unnecessary computation. Further, updates may occur randomly, within dense graph regions --- necessitating processing a substantial portion of the graph. While our earlier work \cite{sahu2024incrementally} had addressed these issues on large dynamic graphs with uniformly random updates, we had observed that our proposed approach did not perform as well on real-world dynamic graphs --- parameter adjustment was needed to achieve acceptable performance. There is thus a need for new approaches that performs well on real-world dynamic graphs, where the nature of updates is different from a uniformly random update.\ignore{In addition, to further improve performance, it is possible to halt rank updates for a vertex if its rank appears to have converged.} This technical report introduces such approaches.

\ignore{In prior research \cite{sahu2024incrementally}, we addressed challenges on large dynamic graphs with random updates but found our approach lacked optimal performance on real-world graphs. Adjusting parameters was necessary for better results. Thus, new approaches tailored for real-world graphs, which experience different update patterns, are introduced in this report.\ignore{Additionally, to enhance efficiency, rank updates for vertices can be halted if their ranks stabilize.}}

\subsection{Our Contributions}

This report presents our improved Dynamic Frontier (DF) and Dynamic Frontier with Pruning (DF-P) approaches\footnote{\url{https://github.com/puzzlef/pagerank-openmp-dynamic}} for updating PageRank on dynamic graphs. These approaches efficiently identify vertices likely to change ranks upon batch updates, with minimal overhead. On a server with a 64-core AMD EPYC-7742 processor, our approaches outperform Static and Dynamic Traversal PageRank by $5.2\times$/$15.2\times$ and $1.3\times$/$3.5\times$ respectively on real-world dynamic graphs, and by $7.2\times$/$9.6\times$ and $4.0\times$/$5.6\times$ on large static graphs with random batch updates. Our observations indicate that the speedup offered by DF and DF-P PageRank mainly stems from the incremental marking of affected vertices. Additionally, our approaches show performance gains of $1.8\times$/$1.7\times$ for every doubling of threads.

\section{Related work}
\label{sec:related}
Early work in dynamic graph algorithms in the sequential setting includes the sparsification method proposed by Eppstein et al. \cite{graph-eppstein97} and Ramalingam's bounded incremental computation approach \cite{incr-ramalingam96}.\ignore{The latter advocates measuring the work done as part of the update in proportion to the effect the update has on the computation.} Several approaches have been suggested for incremental computation of approximate PageRank values in a dynamic or evolving graph. Chien et al. \cite{rank-chien01} identify a small region near updated vertices in the graph and represent the rest of the graph as a single vertex in a smaller graph. PageRanks are computed for this reduced graph and then transferred back to the original graph. Chen et al. \cite{chen2004local} propose various methods to estimate the PageRank score of a webpage using a small subgraph of the entire web, by expanding backwards from the target node along reverse hyperlinks. Bahmani et al. \cite{bahmani2010fast} analyze the efficiency of Monte Carlo methods for incremental PageRank computation. Zhan et al. \cite{zhan2019fast} introduce a Monte Carlo-based algorithm for PageRank tracking on dynamic networks, maintaining $R$ random walks starting from each node. Pashikanti et al. \cite{rank-pashikanti22} also employ a similar Monte Carlo-based approach for updating PageRank scores upon vertex and edge insertions/deletions.

A few approaches have been devised to update exact PageRank scores on dynamic graphs. Zhang \cite{rank-zhang17} introduces a simple incremental PageRank computation system for dynamic graphs, which we refer to as the \textit{Naive-dynamic (ND)} approach, on hybrid CPU and GPU platforms\ignore{ --- employing the Update-Gather-Apply-Scatter (UGAS) computation model}. Additionally, Ohsaka et al. \cite{ohsaka2015efficient} propose a method for locally updating PageRank using the Gauss-Southwell method, prioritizing the vertex with the greatest residual for initial updating; however, their algorithm is inherently sequential. A widely adopted approach for updating PageRank \cite{rank-desikan05, kim2015incremental, rank-giri20, sahu2022dynamic} is based on the observation that changes in the out-degree of a node do not influence its PageRank score, adhering to the first-order Markov property. The portion of the graph undergoing updates, involving edge insertions or deletions, is used to identify the affected region of the graph in a preprocessing step. This is typically accomplished through Breadth-First Search (BFS) or Depth-First Search (DFS) traversal from vertices connected to the inserted or deleted edges. Subsequently, PageRanks are computed solely for this region. Desikan et al. \cite{rank-desikan05} originally proposed this, which we term as the \textit{Dynamic Traversal (DT)} approach in this report. Kim and Choi \cite{kim2015incremental} apply this approach with an asynchronous PageRank implementation, while Giri et al. \cite{rank-giri20} utilize it with collaborative executions on multi-core CPUs and massively parallel GPUs. Sahu et al. \cite{sahu2022dynamic} employ this strategy on a Strongly Connected Component (SCC)-based graph decomposition to limit computation to reachable SCCs from updated vertices, on multi-core CPUs and GPUs.

In our previous study \cite{sahu2024incrementally}, we introduced an incrementally expanding method for updating PageRank on dynamic graphs, demonstrating strong performance on dynamic graphs derived from large static graphs with uniformly random batch updates. However, we noted that the approach did not perform as effectively on real-world dynamic graphs. Adjusting parameters, specifically lowering the frontier tolerance, was necessary to achieve decent performance. Therefore, the selection of frontier tolerance, along with the method of frontier expansion, relies on the nature of batch updates.\ignore{In this technical report, we explore how to expand the frontier, along with the choice of a suitable frontier tolerance, for real-world dynamic graphs. Next, we explore how to prune processed vertices, i.e., contract the set of affected vertices, and select an appropriate prune tolerance. Our experiments on both real-world dynamic graphs and large static graphs with uniformly random batch updates demonstrate the effectiveness of our approach with minor adjustments in both scenarios. Note that a batch update refers to a set of simultaneous changes, i.e., edge insertions or deletions, applied to the graph in a single step.}

Further, Bahmani et al. \cite{rank-bahmani12} introduce an algorithm for selectively crawling a small section of the web to estimate the true PageRank of the graph at a given moment, while Berberich et al. \cite{rank-berberich07} propose a method to compute normalized PageRank scores that remain robust against non-local changes in the graph. These approaches diverge from our improved \textit{Dynamic Frontier} approach, which concentrates on computing the PageRank vector itself rather than on the tasks of web crawling or maintaining normalized scores.

\section{Preliminaries}
\label{sec:preliminaries}
\subsection{PageRank algorithm}
\label{sec:pagerank}

The PageRank, denoted as $R[v]$, of a vertex $v \in V$ in the graph $G(V, E)$, quantifies its \textit{importance} based on the number and significance of incoming links. Equation \ref{eq:pr} outlines the computation of PageRank for vertex $v$ in graph $G$, where $V$ represents the set of vertices\ignore{($N = |V|$)}, $E$ represents the set of edges\ignore{($M = |E|$)}, $G.in(v)$ denotes the incoming neighbors of vertex $v$, $G.out(v)$ denotes the outgoing neighbors of vertex $v$, and $\alpha$ represents the damping factor. Initially, each vertex has a PageRank of $1/|V|$. The \textit{power-iteration} method iteratively updates these values until they converge within a specified tolerance $\tau$. This is typically measured using the $L_1$-norm \cite{ohsaka2015efficient}, though $L_2$ and $L_\infty$-norm are also occasionally used.

The random surfer model, integral to the PageRank algorithm, conceptualizes a surfer navigating the web by following links on each page. The damping factor $\alpha$, with a default value of $0.85$, represents the probability that the surfer continues along a link instead of jumping randomly. PageRank for each page reflects the long-term likelihood of the surfer visiting that page, based on starting from a random page and following links\ignore{according to the damping factor}. PageRank values are essentially the eigenvector of a transition matrix, which encodes probabilities of moving between pages in a Markov Chain.

Dead ends, also known as dangling vertices, pose a challenge in PageRank computation. They are vertices with no out-links, ans thus force the surfer to jump to a random web page. Consequently, dead ends contribute their rank equally among all vertices in the graph --- this must be computed in each iteration, and is therefore an overhead. We address this issue by adding self-loops to all vertices in the graph \cite{kolda2009generalized, rank-andersen07, rank-langville06}. In a streaming environment, this option may be the most suitable. It has also been observed to be superior in spam-link applications \cite{kolda2009generalized}.

\begin{equation}
\label{eq:pr}
    R[v] = \alpha \times \sum_{u \in G.in(v)} \frac{R[u]}{|G.out(u)|} + \frac{1 - \alpha}{|V|}
\end{equation}

\subsection{Dynamic Graphs}
\label{sec:about-dynamic}

A dynamic graph can be conceptualized as a sequence of graphs, where $G^t(V^t, E^t)$ represents the graph at time step $t$. The changes between consecutive time steps $t-1$ and $t$, from $G^{t-1}(V^{t-1}, E^{t-1})$ to $G^t(V^t, E^t)$, can be represented as a batch update $\Delta^t$ at time step $t$. This update comprises a set of edge deletions $\Delta^{t-}$, defined as $\{(u, v)\ |\ u, v \in V\} = E^{t-1} \setminus E^t$, and a set of edge insertions $\Delta^{t+}$, defined as $\{(u, v)\ |\ u, v \in V\} = E^t \setminus E^{t-1}$.

\paragraph{Interleaving graph updates with computation:}

We assume changes to the graph to be batched, with updating of the graph and algorithm execution occurring in an interleaved manner --- allowing only one writer on the graph structure at any given time. If it is needed to update the graph in parallel with the computation, a graph snapshot needs to be obtained, on which the computation can be performed. See for example, the Aspen graph processing framework, which minimizes\ignore{read-only} snapshot acquisition costs \cite{graph-dhulipala19}.

\subsection{Existing approaches for updating PageRank on Dynamic Graphs}

\subsubsection{Naive-dynamic approach}
\label{sec:about-naive}

This\ignore{straightforward} approach involves updating vertex ranks in dynamic networks by initializing them with ranks from the previous graph snapshot and running the PageRank algorithm on all vertices. Rankings obtained using this approach are at least as accurate as those obtained from the static algorithm.\ignore{Zhang et al. \cite{rank-zhang17} have explored the \textit{Naive-dynamic} approach in the hybrid CPU-GPU setting.}

\subsubsection{Dynamic Traversal approach}
\label{sec:about-traversal}

Initially proposed by Desikan et al. \cite{rank-desikan05}, this approach involves skipping the processing of vertices whose ranks cannot be impacted by the given batch update. For each edge deletion or insertion $(u, v)$ in the batch update, all vertices reachable from vertex $u$ in either graph $G^{t-1}$ or $G^t$ are marked as affected, using DFS or BFS.\ignore{Giri et al. \cite{rank-giri20} have explored the \textit{Dynamic Traversal} approach in the hybrid CPU-GPU setting. On the other hand, Banerjee et al. \cite{rank-sahu22} have explored this approach in the CPU and GPU settings separately where they compute the ranks of vertices in topological order of strongly connected components (SCCs) to minimize unnecessary computation. They borrow this ordered processing of SCCs from the original static algorithm proposed by Garg et al. \cite{rank-garg16}.}

\section{Approach}
\label{sec:approach}
In the event of a batch update $\Delta^{t-} \cup \Delta^{t+}$ being relatively small compared to the total number of edges $|E|$, it is expected that only a small subset of vertices will undergo rank changes. To tackle this situation, our proposed approaches utilize an incremental process to identify affected vertices and update their ranks.

\begin{figure*}[hbtp]
  \centering
  \subfigure[Initial graph]{
    \label{fig:about-frontier-df1}
    \includegraphics[width=0.23\linewidth]{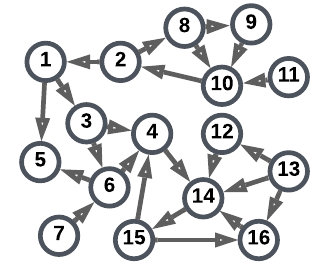}
  }
  \subfigure[Marking initial affected vertices (DF)]{
    \label{fig:about-frontier-df2}
    \includegraphics[width=0.23\linewidth]{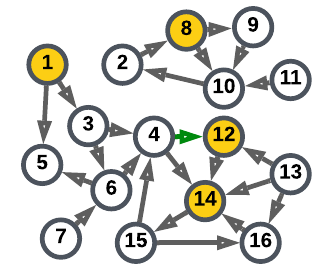}
  }
  \subfigure[After first iteration (DF)]{
    \label{fig:about-frontier-df3}
    \includegraphics[width=0.23\linewidth]{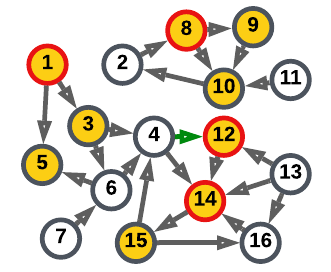}
  }
  \subfigure[After second iteration (DF)]{
    \label{fig:about-frontier-df4}
    \includegraphics[width=0.23\linewidth]{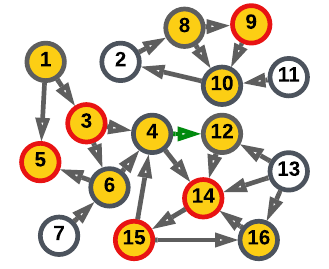}
  } \\[2ex]
  \subfigure[Initial graph]{
    \label{fig:about-frontier-dfp1}
    \includegraphics[width=0.23\linewidth]{out/about-frontier-11.pdf}
  }
  \subfigure[Marking initial affected vertices (DF-P)]{
    \label{fig:about-frontier-dfp2}
    \includegraphics[width=0.23\linewidth]{out/about-frontier-32.pdf}
  }
  \subfigure[After first iteration (DF-P)]{
    \label{fig:about-frontier-dfp3}
    \includegraphics[width=0.23\linewidth]{out/about-frontier-33.pdf}
  }
  \subfigure[After second iteration (DF-P)]{
    \label{fig:about-frontier-dfp4}
    \includegraphics[width=0.23\linewidth]{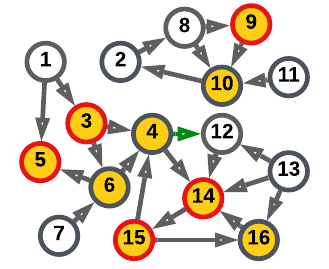}
  } \\[2ex]
  \subfigure[Initial graph]{
    \label{fig:about-frontier-dt1}
    \includegraphics[width=0.23\linewidth]{out/about-frontier-11.pdf}
  }
  \subfigure[Marking affected vertices (DT)]{
    \label{fig:about-frontier-dt2}
    \includegraphics[width=0.23\linewidth]{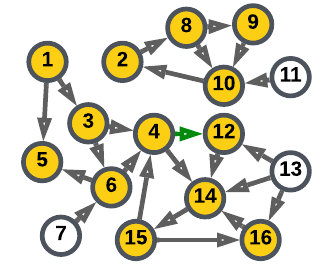}
  }
  \subfigure[After first iteration (DT)]{
    \label{fig:about-frontier-dt3}
    \includegraphics[width=0.23\linewidth]{out/about-frontier-22.pdf}
  }
  \subfigure[After second iteration (DT)]{
    \label{fig:about-frontier-dt4}
    \includegraphics[width=0.23\linewidth]{out/about-frontier-22.pdf}
  } \\[-2ex]
  \caption{An example showcasing our improved \textit{Dynamic Frontier (DF)} and \textit{Dynamic Frontier with Pruning (DF-P)} approaches, in subfigures (a)-(d) and (e)-(h) respectively, in contrast to the \textit{Dynamic Traversal (DT)} approach, shown in subfigures (i)-(l).}
  \label{fig:about-frontier}
\end{figure*}

\subsection{Our improved Dynamic Frontier approaches}
\label{sec:frontier}

We now explain our improved \textit{Dynamic Frontier (DF)} and \textit{Dynamic Frontier with Pruning (DF-P)} approaches. Consider a batch update with edge deletions $(u, v) \in \Delta^{t-}$ and insertions $(u, v) \in \Delta^{t+}$.

\subsubsection{Our improved Dynamic Frontier (DF) PageRank}

\paragraph{Initialization of ranks:}

Initially, we set the rank of each vertex to match the rank it had in the previous snapshot of the graph.

\paragraph{Initial marking of affected vertices:}

For every edge deletion/insertion $(u, v)$, mark the outgoing neighbors of vertex $u$ in both the previous snapshot $G^{t-1}$ and the current graph snapshot $G^t$, as affected.

\paragraph{Incremental expansion of the set of affected vertices upon change in rank of a given vertex:}

During the PageRank computation, if the rank of any affected vertex $v$ changes by a fraction exceeding the \textit{frontier tolerance} $\tau_f$, we designate its outgoing neighbors as affected. This step is taken because a modification in a vertex's rank is likely to impact the ranks of its outgoing neighbors. This process of marking vertices as affected continues in every iteration until the ranks have converged, as indicated by the iteration tolerance $\tau$.

\subsubsection{Our Dynamic Frontier with Pruning (DF-P) PageRank}

\paragraph{Initialization of ranks:}

We set the rank of each vertex to match the rank it had in the preceding snapshot of the graph.

\paragraph{Initial marking of affected vertices:}

For each edge deletion/insertion $(u, v)$, we mark the outgoing neighbors of vertex $u$ in both the previous $G^{t-1}$ and the current snapshot $G^t$ of the graph as affected.

\paragraph{Incremental expansion and contraction of the set of affected vertices upon change in rank of a given vertex:}

During PageRank computation, if the rank of any affected vertex $v$ changes in an iteration by a fraction greater than the \textit{frontier tolerance} $\tau_f$, we mark its outgoing neighbors as affected. Additionally, if the relative change in rank of a vertex remains below the \textit{prune tolerance} $\tau_p$, indicating potential convergence, the vertex is no longer marked as affected. However, if its rank has not converged, it may be re-marked as affected by one of its in-neighbors. This marking and unmarking process continues in every iteration, until the ranks have converged.

\paragraph{Computation of rank of each vertex}

As each vertex may be pruned (or unmarked as affected), and given that each vertex has a self-loop (as described in Sections \ref{sec:dataset} and \ref{sec:batch-generation}), we employ a closed-loop formula to calculate the rank of each vertex (Equation \ref{eq:pr-prune}). This formula accounts for the self-loop's presence, thereby reducing the need for recursive rank calculation due to the self-loop. The derivation of this formula is detailed in Section \ref{sec:pr-prune-derivation}.

\begin{flalign}
\label{eq:pr-prune}
  R[v] & = \frac{1}{1 - \alpha / |G.out(v)|} \left(\alpha K + \frac{1 - \alpha}{|V|}\right) && \\
    \text{where, } K & = \left(\sum_{u \in G.in(v)} \frac{R[u]}{|G.out(u)|}\right) - \frac{R[v]}{|G.out(v)|}
\end{flalign}

\subsubsection{A simple example}

Figure \ref{fig:about-frontier} illustrates an example of our improved Dynamic Frontier (DF) and Dynamic Frontier with Pruning (DF-P) PageRank. Initially, as depicted in Figures \ref{fig:about-frontier-df1} and \ref{fig:about-frontier-dfp1}, the graph comprises $16$ vertices and $23$ edges. Subsequently, Figures \ref{fig:about-frontier-df2} and \ref{fig:about-frontier-dfp2} show a batch update applied to the original graph, involving an edge insertion from vertex $4$ to $12$ and an edge deletion from vertex $2$ to $1$. Following the batch update, we proceed with the initial step of DF/DF-P PageRank, marking the outgoing neighbors of vertices $2$ and $4$ as affected, specifically vertices $1$, $8$, $12$, and $14$. These affected vertices are highlighted with a yellow fill. It may be noted that vertices $2$ and $4$ are not marked as affected. This is because changes in the out-degree of a vertex does not influence its PageRank score (see Equation \ref{eq:pr}). Subsequently, we initiate the first iteration of the PageRank algorithm.

During the first iteration (refer to Figures \ref{fig:about-frontier-df3} and \ref{fig:about-frontier-dfp3}), the ranks of affected vertices are updated. It is observed that the relative change in rank of vertices $1$, $8$, $12$, and $14$ exceeds the frontier tolerance $\tau_f$. Such vertices are indicated with a red border in the figures. In response to this, with both DF and DF-P PageRank, we incrementally mark the outgoing neighbors of vertices $1$, $8$, $12$, and $14$ as affected, specifically vertices $3$, $5$, $9$, $10$, $14$, and $15$

In the second iteration, shown in Figures \ref{fig:about-frontier-df4} and \ref{fig:about-frontier-dfp4}, updates are made to the ranks of affected vertices once again. Here, it is observed that the relative change in rank of vertices $3$, $5$, $9$, $14$, and $15$ exceeds the frontier tolerance $\tau_f$. Consequently, with DF/DF-P PageRank, we mark the outgoing neighbors of vertices $3$, $5$, $9$, $14$, and $15$ as affected, specifically vertices $4$, $6$, $10$, $15$, and $16$. Moreover, it is observed that the relative change in rank of vertices $1$, $8$, and $12$ remains below the prune tolerance $\tau_p$. As a result, with DF-P PageRank, these vertices are no longer marked as affected, as it is likely the ranks of such vertices have converged. This action effectively contracts the frontier of affected vertices. However, if the rank of such a vertex has not yet converged, it may be re-marked as affected by one of its in-neighbors.

In the next iteration, the ranks of affected vertices are updated once more. If the change in rank of each vertex remains within the iteration tolerance $\tau$ (we use $L\infty$-norm for convergence detection), the ranks of vertices have converged, and the algorithm terminates.

\paragraph{Contrasting with Dynamic Traversal (DT) PageRank}

Let us now contrast with DF and DF-P PageRank with Dynamic Traversal (DT) PageRank, shown in Figures \ref{fig:about-frontier-dt1}-\ref{fig:about-frontier-dt4}. Figure \ref{fig:about-frontier-dt2} show the same batch update applied to the original graph, as in Figures \ref{fig:about-frontier-df2} and \ref{fig:about-frontier-dfp2}. In response to this, DT PageRank marks all vertices reachable from $2$ and $4$ as affected, i.e., all vertices except $7$, $11$, and $13$. The ranks of this set of affected vertices are then updated in each iteration (ranks of unaffected vertices cannot change), until convergence.

\subsection{Determination of Frontier tolerance ($\tau_f$)}
\label{sec:frontier-tolerance}

We first need to determine a suitable approach for frontier expansion, and an associate frontier tolerance $\tau_f$ value that allows us to minimize processed vertices, while limiting error to that of ranks obtained with Static PageRank using the same iteration tolerance $\tau$. For this, we experiment with three approaches. These include marking neighbors of a vertex as affected, based on change in rank of the vertex $\Delta r$, change in its contribution factor $\Delta r/d$, or relative change in its rank $\Delta r/r$. Here, $\Delta r$ is the rank change, $d$ is the out-degree, and $r$ is the \texttt{max} of its previous and current rank values.

\begin{figure*}[!hbt]
  \centering
  \subfigure[Speedup with varying Frontier tolerance $\tau_f$]{
    \label{fig:adjust-frontier--speedup}
    \includegraphics[width=0.48\linewidth]{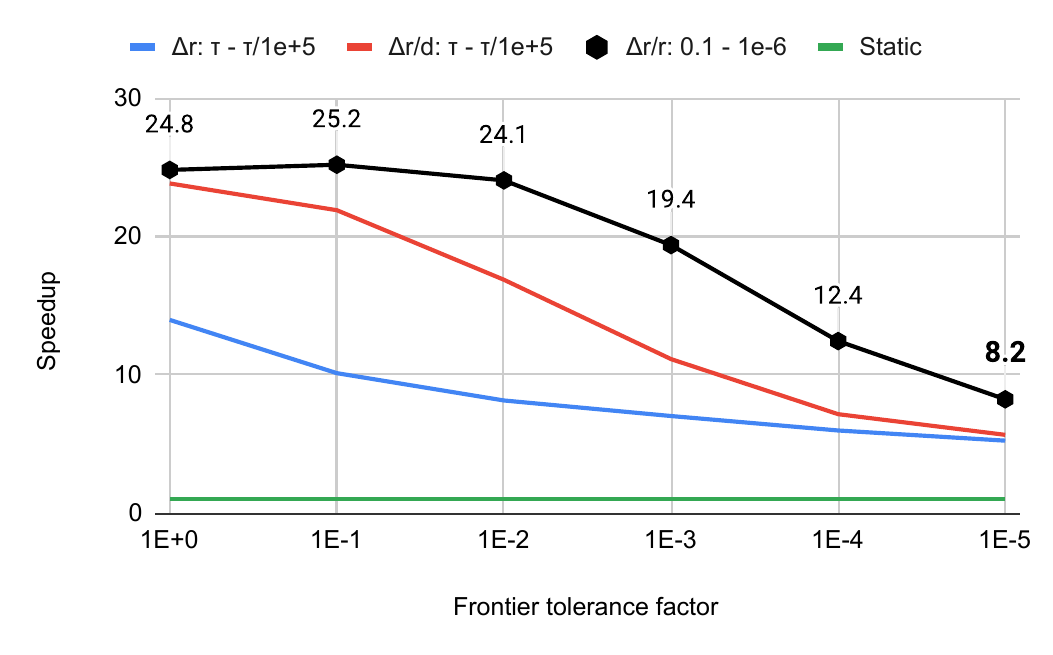}
  }
  \subfigure[Error in ranks obtained with varying Frontier tolerance $\tau_f$]{
    \label{fig:adjust-frontier--error}
    \includegraphics[width=0.48\linewidth]{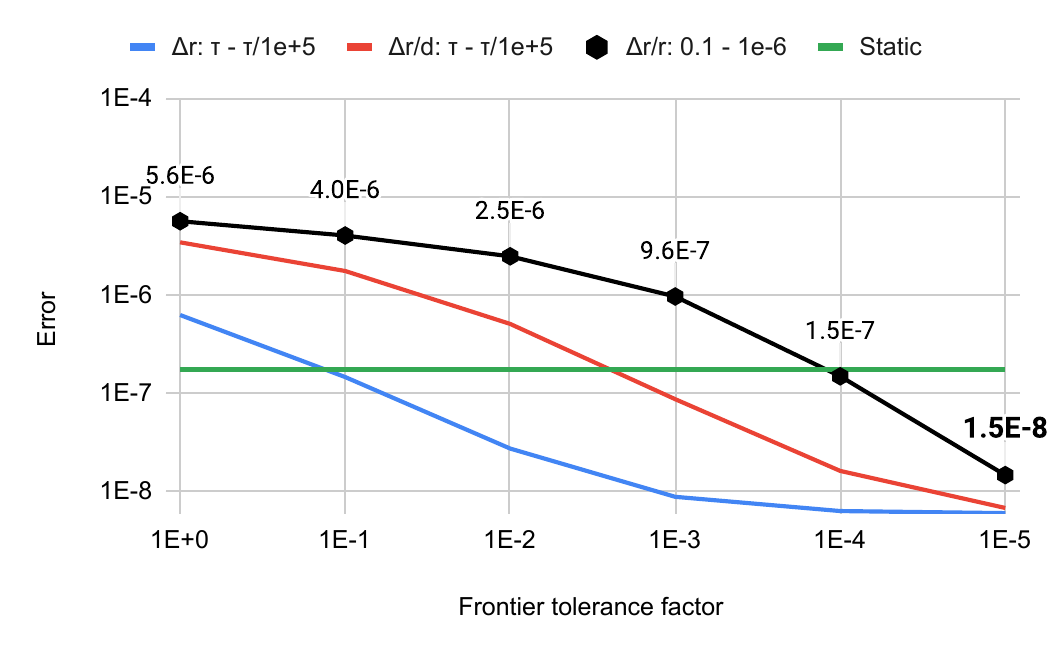}
  } \\[-2ex]
  \caption{Mean Speedup and Error in ranks obtained with three different frontier expansion approaches: Change in rank ($\Delta r$), Change in contribution factor ($\Delta r/d$), and Relative change in rank ($\Delta r/r$). Here, $\Delta r$ represents the change in rank of a vertex, $d$ is its out-degree, and $r$ is the maximum of the previous and current rank value of the vertex. For the first two approaches, we adjust the frontier tolerance $\tau_f$ from $\tau$ to $\tau/10^5$ ($\tau$ is iteration tolerance), and for the last approach, we adjust it from $0.1$ to $10^{-6}$. With each approach, we mark outgoing neighbors as affected if the defined metric exceeds $\tau_f$. We also include the mean speedup and error in ranks obtained with Static PageRank as a reference. This figure demonstrates that the $\Delta r/r$ approach with a $\tau_f$ of $10^{-6}$ performs the best, while achieving ranks with lower error than Static PageRank.}
  \label{fig:adjust-frontier}
\end{figure*}

For $\Delta r$ and $\Delta r/d$, we adjust $\tau_f$ from $\tau$ to $\tau/10^5$; and for $\Delta r/r$, we adjust it from $0.1$ to $10^{-6}$. This is done on real-world dynamic graphs, shown in Table \ref{tab:dataset}, with batch updates of size $10^{-5}|E_T|$. Outgoing neighbors are marked affected if the respective measure exceeds $\tau_f$. Figure \ref{fig:adjust-frontier} shows the mean speedup (with respect to Static PageRank) and rank error (compared to ranks obtained with reference Static PageRank) with each approach for frontier expansion. Results indicate that the $\Delta r/r$ approach with a $\tau_f$ of $10^{-6}$ performs best, while yielding lower error than Static PageRank.

\ignore{Mean Speedup and Error in ranks obtained (with respect to ranks obtained with Reference Static PageRank) with three different approaches of expanding frontier, i.e., marking outgoing neighbors as affected, based on: Change in rank $\Delta r$, Change in contribution factor $\Delta r/d$, and Relative change in rank $\Delta r/r$ of a vertex. Here, $\Delta r$ is the change in rank of a vertex, $d$ is its out-degree, and $r$ represents the rank of the vertex (it is actually the \texttt{max} of the previous and the current rank value of the vertex). For the first two approaches for expanding frontier, i.e., $\Delta r$ and $\Delta r/d$, we adjust the frontier tolerance $\tau_f$ from $\tau$ to $\tau/10^5$ (where $\tau$ is the iteration tolerance), and for the last approach, i.e., $\Delta r/r$, we adjust the frontier tolerance from $0.1$ to $10^-6$ (note that the $0.1$ beside $\Delta r/r$ indicates that the frontier tolerance $\tau_f$ starts from $0.1$). With each approach, we mark the outgoing neighbors of a vertex as affected, only if the metric/measure defined by the approach exceeds the frontier tolerance $\tau_f$. We also include the mean speedup ($1$) and error in ranks obtained ($1.7\times10^{-7}$) with Static PageRank as a reference. This figure indicates that the $\Delta r/r$ approach with a frontier tolerance $\tau_f$ of $10^{-6}$ (i.e., frontier tolerance factor of $10^{-5}$) performs the best, while obtaining ranks with lower error than Static PageRank.}

\begin{figure*}[!hbt]
  \centering
  \subfigure[Speedup with varying Prune tolerance $\tau_p$]{
    \label{fig:adjust-prune--speedup}
    \includegraphics[width=0.48\linewidth]{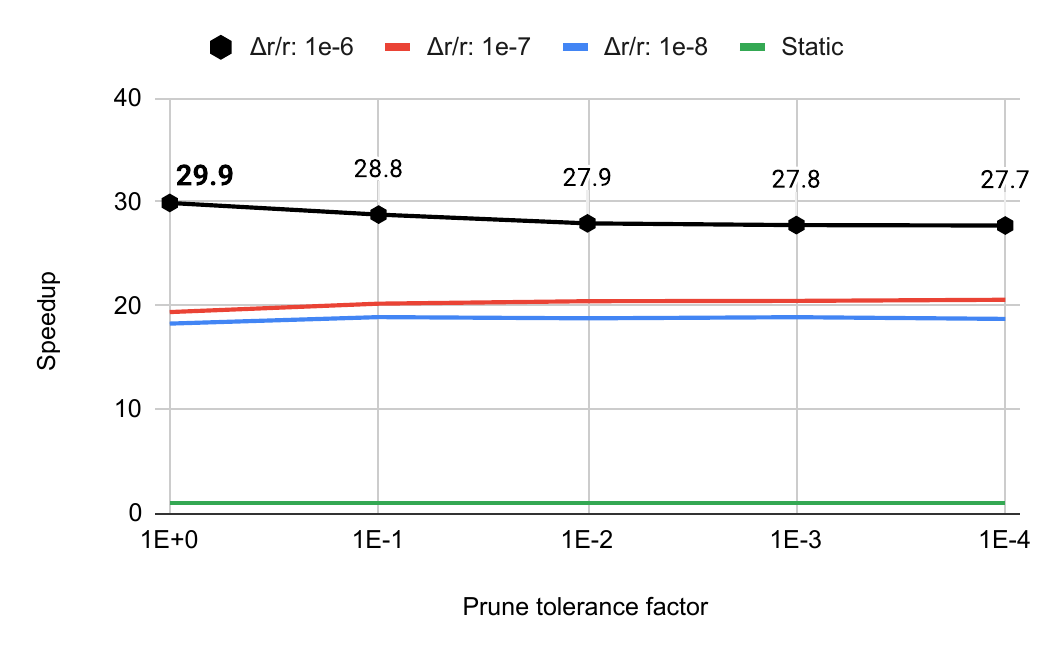}
  }
  \subfigure[Error in ranks obtained with varying Prune tolerance $\tau_p$]{
    \label{fig:adjust-prune--error}
    \includegraphics[width=0.48\linewidth]{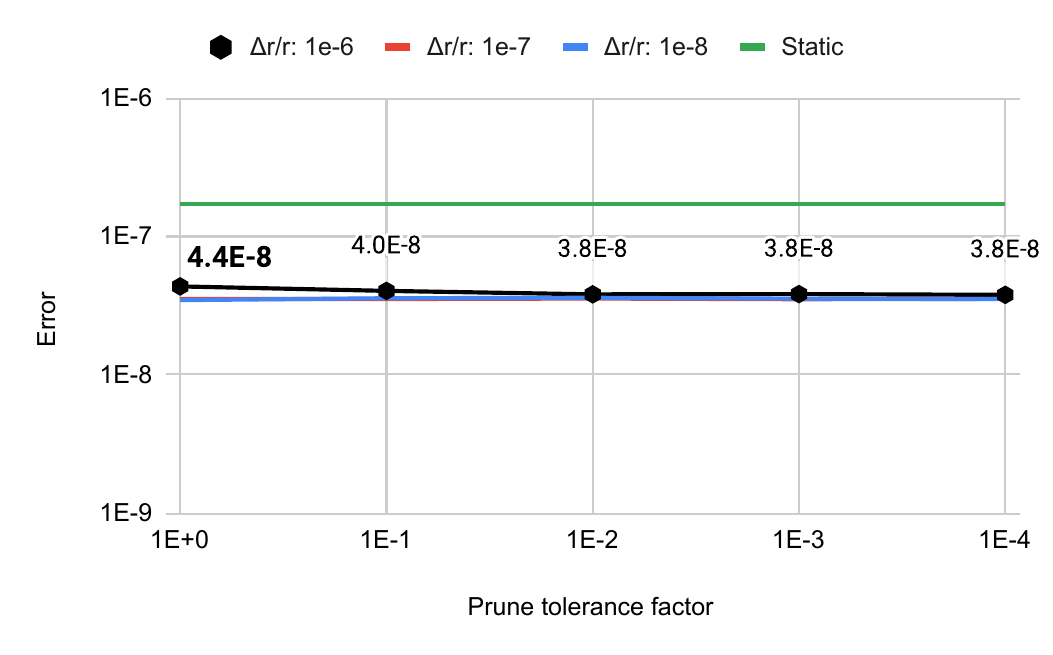}
  } \\[-2ex]
  \caption{Mean Speedup and Error in ranks obtained, with varying prune tolerance $\tau_p$ from $\tau_f$ to $\tau_f/10^4$ ($\tau_f$ is frontier tolerance), using the optimal approach of expanding frontier, i.e., based on relative change in rank $\Delta r/r$ of a vertex with a $\tau_f$ of $10^{-6}$. In addition to a $\tau_f$ of $10^{-6}$, we experiment with $\tau_f$ of $10^{-7}$ and $10^{-8}$. We also plot Static PageRank as a reference. The $\Delta r/r$ approach with a $\tau_f$ of $10^{-6}$ and a $\tau_p$ of $\tau_f$ performs the best, with lower error than Static PageRank.}
  \label{fig:adjust-prune}
\end{figure*}

\ignore{Mean Speedup and Error in ranks obtained (with respect to ranks obtained with Reference Static PageRank) with the best approach of expanding frontier, i.e., marking outgoing neighbors as affected, based on relative change in rank $\Delta r/r$ of a vertex, with a frontier tolerance $\tau_f$ of $10^{-6}$. In the figure, we vary prune tolerance $\tau_p$ from $\tau_f$ to $\tau_f/10^5$. When the relative change in rank of a vertex falls within prune tolerance $\tau_p$, we mark the vertex as not affected. In addition to a frontier tolerance $\tau_f$ of $10^{-6}$, we also experiment with a $\tau_f$ of $10^{-7}$ and $10^{-8}$. We also include the mean speedup ($1$) and error in ranks obtained ($1.7\times10^{-7}$) with Static PageRank as a reference. Here, $\Delta r$ is the change in rank of a vertex and $r$ represents the rank of the vertex (it is actually the \texttt{max} of the previous and the current rank value of the vertex). This figure indicates that the $\Delta r/r$ approach with a frontier tolerance $\tau_f$ of $10^{-6}$ and a prune tolerance $\tau_p$ of $\tau_f = 10^{-6}$ performs the best, while obtaining ranks with lower error than Static PageRank.}

\begin{algorithm}[!hbt]
\caption{Our parallel Dynamic Frontier (DF*) PageRank.}
\label{alg:frontier}
\begin{algorithmic}[1]
\Require{$G^{t-1}, G^t$: Previous, current input graph}
\Require{$\Delta^{t-}, \Delta^{t+}$: Edge deletions and insertions (input)}
\Require{$R^{t-1}, R$: Previous, current rank vector}
\Ensure{$\Delta r$: Change in rank of a vertex}
\Ensure{$\Delta R$: $L\infty$-norm between previous and current ranks}
\Ensure{$\tau, \tau_f, \tau_p$: Iteration, frontier, prune tolerance}
\Ensure{$\alpha$: Damping factor}

\Statex

\Function{dynamicFrontier}{$G^{t-1}, G^t, \Delta^{t-}, \Delta^{t+}, R^{t-1}$}
  \State $R \gets R^{t-1}$ \label{alg:frontier--initialize}
  \State $\rhd$ Mark initial affected
  \ForAll{$(u, v) \in \Delta^{t-} \cup \Delta^{t+} \textbf{in parallel}$} \label{alg:frontier--mark-begin}
    \ForAll{$v' \in (G^{t-1} \cup G^t).out(u)$}
    \State Mark $v'$ as affected
    \EndFor
  \EndFor \label{alg:frontier--mark-end}
  \ForAll{$i \in [0 .. MAX\_ITERATIONS)$} \label{alg:frontier--compute-begin}
    \State $\Delta R \gets 0$ \textbf{;} $C_0 \gets (1 - \alpha)/|V^t|$
    \ForAll{affected $v \in V^t$ \textbf{in parallel}}
      \State $c \gets 0$ \textbf{;} $d \gets |G^t.out(v)|$
      \ForAll{$u \in G^t.in(v)$}
        \State $c \gets c + R[u] / |G^t.out(u)|$
      \EndFor
      \If{\textbf{is \textit{DF-P}}} \label{alg:frontier--formula-begin}
        \State $r \gets 1/(1 - \alpha/d) * (C_0 + \alpha * (c - R[v]/d))$
      \Else
        \State $r \gets C_0 + \alpha * c$
      \EndIf \label{alg:frontier--formula-end}
      \State $\Delta r \gets |r - R[v]|$ \textbf{;} $\Delta R \gets \max(\Delta R, \Delta r)$
      \State $\rhd$ Prune $v$ if its relative rank change is small
      \If{\textbf{is \textit{DF-P} and} $\Delta r / \max(r, R[v]) \leq \tau_p$}
        \State Mark $v$ as not affected
      \EndIf
      \State $\rhd$ Expand frontier if relative rank change is large
      \If{$\Delta r / \max(r, R[v]) > \tau_f$} \label{alg:frontier--remark-begin}
        \ForAll{$v' \in G^t.out(v)$}
          \State Mark $v'$ as affected
        \EndFor
      \EndIf \label{alg:frontier--remark-end}
      \State $\rhd$ Update rank of $v$
      \State $R[v] \gets r$
    \EndFor
    \State $\rhd$ Ranks converged?
    \If{$\Delta R \le \tau$} \textbf{break}
    \EndIf
  \EndFor \label{alg:frontier--compute-end}
  \State \ReturnInline{$R$} \label{alg:frontier--return}
\EndFunction
\end{algorithmic}
\end{algorithm}



\subsection{Determination of Prune tolerance ($\tau_p$)}
\label{sec:prune-tolerance}

We now embark on determining a suitable value for the prune tolerance $\tau_p$ to complement the optimal frontier expansion approach $\Delta r/r$, which employs a frontier tolerance $\tau_f$ of $10^{-6}$ as identified in Section \ref{sec:frontier-tolerance}. This entails adjusting $\tau_p$ from $\tau_f$ to $\tau_f/10^4$. Additionally, to err on the side of caution, we explore the effects of lower $\tau_f$ values, namely $10^{-7}$ and $10^{-8}$. These experiments are conducted on real-world graphs, employing batch updates of size $10^{-5}|E_T|$ as outlined earlier. A vertex is categorized as unaffected if its relative rank change $\Delta r/r$ falls within the designated $\tau_p$ range.

Figure \ref{fig:adjust-prune} presents the mean speedup, compared to Static PageRank, and the corresponding rank error observed when employing different $\tau_f$ values for frontier expansion. The rank error is measured with respect to reference Static PageRank, as discussed in Section \ref{sec:measurement}. Notably, the results highlight that the $\Delta r/r$ approach, particularly with a $\tau_f$ set to $10^{-6}$ and an accompanying $\tau_p$ of $\tau_f = 10^{-6}$, achieves superior performance by attaining lower rank error compared to Static PageRank.

\subsection{Our DF* PageRank implementation}

Algorithm \ref{alg:frontier} shows the pseudocode of our improved Dynamic Frontier (DF) and Dynamic Frontier with Pruning (DF-P) PageRank. It takes as input the previous $G^{t-1}$ and current $G^t$ snapshot of the graph, edge deletions $\Delta^{t-}$ and insertions $\Delta^{t+}$ in the batch update, the previous rank vector $R^{t-1}$, and returns the updated ranks $R$.

The algorithm begins by initializing the current rank vector $R$ with the previous rank vector $R^{t-1}$ (line \ref{alg:frontier--initialize}), and marking the initially affected vertices based on edge deletions $\Delta^{t-}$ and insertions $\Delta^{t+}$ in parallel (lines \ref{alg:frontier--mark-begin}-\ref{alg:frontier--mark-end}). It then iteratively computes the rank $R[v]$ for each affected vertex $v$ (lines \ref{alg:frontier--compute-begin}-\ref{alg:frontier--compute-end}). This computation is performed in parallel, considering the incoming edges $G^t.in(v)$. Depending on whether DF or DF-P PageRank is selected, the corresponding formula for rank calculation is applied (lines \ref{alg:frontier--formula-begin}-\ref{alg:frontier--formula-end}). The algorithm then checks if the relative change in rank $\Delta r / \max(r, R[v])$ exceeds the frontier tolerance $\tau_f$, marking out-neighbor vertices as affected if so. Additionally, with DF-P PageRank, if the relative change in rank lies within the prune tolerance $\tau_p$, the vertex $v$ is marked as not affected. The iteration continues until either the maximum change in ranks $\Delta R$ falls below the iteration tolerance $\tau$, or the maximum number of iterations $MAX\_ITERATIONS$ is reached. Finally, the algorithm returns the final rank vector $R$ (line \ref{alg:frontier--return}).

In a push-based approach for PageRank computation, each thread calculates and sums the outgoing PageRank contribution of its vertex to its neighbors, necessitating atomic updates. In contrast, with a pull-based approach, each vertex's rank is updated through a single write by a thread \cite{verstraaten2015quantifying}. We find this to be more efficient and employ it for all implementations. Furthermore, we employ an asynchronous implementation of DF and DF-P PageRank, using a single rank vector, for potentially faster convergence and elimination of memory copies for unaffected vertices. This, based on our previous research \cite{sahu2024incrementally}, outperforms synchronous implementations, especially with smaller batch sizes. We also utilize asynchronous implementations for Naive-dynamic (ND) and Dynamic Traversal (DT) PageRank, but not for Static PageRank (async not faster).

\section{Evaluation}
\label{sec:evaluation}
\subsection{Experimental Setup}
\label{sec:setup}

\subsubsection{System used}

Experiments are performed on a system featuring an AMD EPYC-7742 processor with $64$ cores, operating at a frequency of $2.25$ GHz. Each core is equipped with a $4$ MB L1 cache, a $32$ MB L2 cache, and shares a $256$ MB L3 cache. The server is set up with $512$ GB of DDR4 system memory and runs Ubuntu $20.04$.

\subsubsection{Configuration}

We use 32-bit integers for vertex IDs and 64-bit floating-point numbers for vertex ranks. Affected vertices are represented with an 8-bit integer vector. Rank computation employs OpenMP's \textit{dynamic schedule} with a chunk size of $2048$ for dynamic workload balancing among threads. We set the damping factor to $\alpha = 0.85$ \cite{rank-langville06} and an iteration tolerance of $\tau = 10^{-10}$ using the $L_\infty$-norm \cite{rank-dubey22, rank-plimpton11}. The maximum number of iterations $MAX\_ITERATIONS$ is limited to $500$ \cite{nvgraph}. All experiments run with $64$ threads to match the available system cores, unless stated otherwise. Compilation is done using GCC $9.4$ and OpenMP $5.0$.

\subsubsection{Dataset}
\label{sec:dataset}

We utilize five temporal networks from the Stanford Large Network Dataset Collection \cite{snapnets}, outlined in Table \ref{tab:dataset}. These networks contain vertex counts ranging from $24.8$ thousand to $2.60$ million, temporal edge counts from $507$ thousand to $63.4$ million, and static edge counts from $240$ thousand to $36.2$ million. To address dead ends (vertices lacking out-links), a global teleport rank computation is needed in each iteration. We mitigate this overhead by adding self-loops to all vertices\ignore{ in the graph} \cite{kolda2009generalized, rank-andersen07, rank-langville06}.

\begin{table}[hbtp]
  \centering
  \caption{List of 5 real-world dynamic graphs\ignore{, i.e., temporal networks}, obtained from the Stanford Large Network Dataset Collection \cite{snapnets}. Here, $|V|$ is the number of vertices, $|E_T|$ the number of temporal edges\ignore{(includes duplicate edges)}, and $|E|$ the number of static edges (with no duplicates).\ignore{, and $\Gamma_G$ is the Gini coefficient of PageRank distribution. In the table, B refers to a billion, M refers to a million and K refers a thousand.}}
  \label{tab:dataset}
  \begin{tabular}{|c||c|c|c|c|}
    \toprule
    \textbf{Graph} &
    \textbf{\textbf{$|V|$}} &
    \textbf{\textbf{$|E_T|$}} &
    \textbf{\textbf{$|E|$}} \\
    \midrule
    sx-mathoverflow & 24.8K & 507K & 240K \\ \hline
    sx-askubuntu & 159K & 964K & 597K \\ \hline
    sx-superuser & 194K & 1.44M & 925K \\ \hline
    wiki-talk-temporal & 1.14M & 7.83M & 3.31M \\ \hline
    sx-stackoverflow & 2.60M & 63.4M & 36.2M \\ \hline
  \bottomrule
  \end{tabular}
\end{table}

\subsubsection{Batch Generation}
\label{sec:batch-generation}

In each experiment, we initially load $90\%$ of every real-world dynamic graph from Table \ref{tab:dataset}, followed by loading $B$ edges consecutively in $100$ batch updates. Here, $B$ represents the desired batch size, specified as a fraction of the total number of temporal edges $|E_T|$ in the graph. Additionally, self-loops are added to all vertices with each batch update.

\subsubsection{Measurement}
\label{sec:measurement}

We evaluate the runtime of each approach on the entire updated graph, including preprocessing and convergence detection time, but excluding memory allocation/deallocation time. The mean time and error for a specific method at a given batch size is computed as the geometric mean across input graphs.\ignore{Average speedup is the ratio of these mean times.} Additionally, we assess the error/accuracy of each approach by measuring the $L1$-norm \cite{ohsaka2015efficient} of the ranks compared to ranks obtained from a reference Static PageRank run on the updated graph with an extremely low iteration tolerance of $\tau = 10^{-100}$ (limited to $500$ iterations).

\subsection{Performance comparison}

\subsubsection{Results on real-world dynamic graphs}

\begin{figure*}[!hbt]
  \centering
  \subfigure[Overall Runtime]{
    \label{fig:temporal-summary--runtime-overall}
    \includegraphics[width=0.48\linewidth]{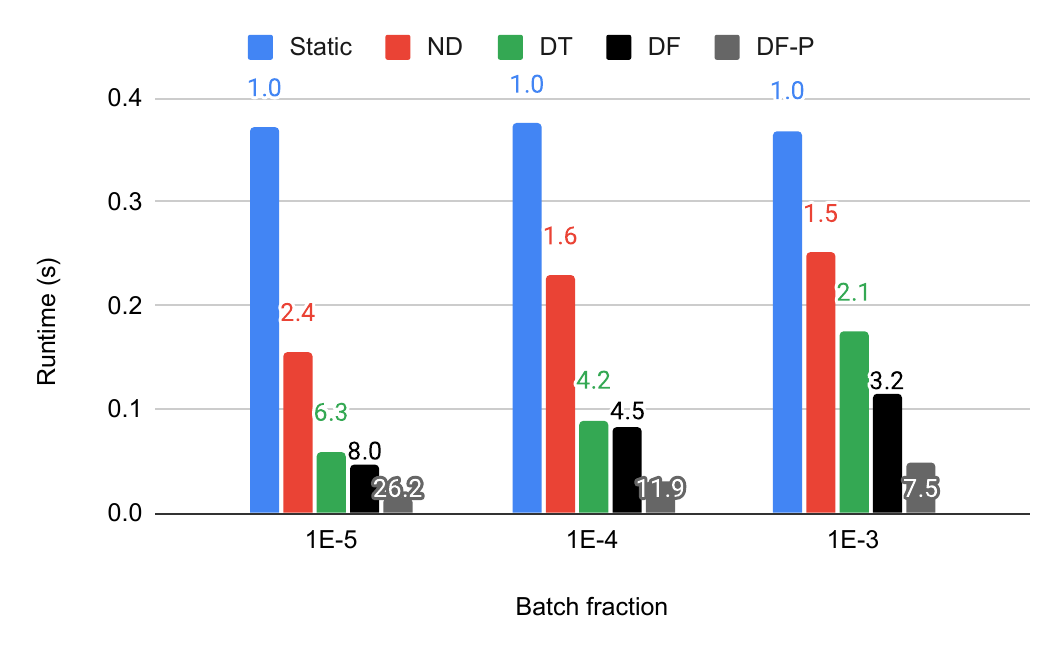}
  }
  \subfigure[Overall Error in ranks obtained]{
    \label{fig:temporal-summary--error-overall}
    \includegraphics[width=0.48\linewidth]{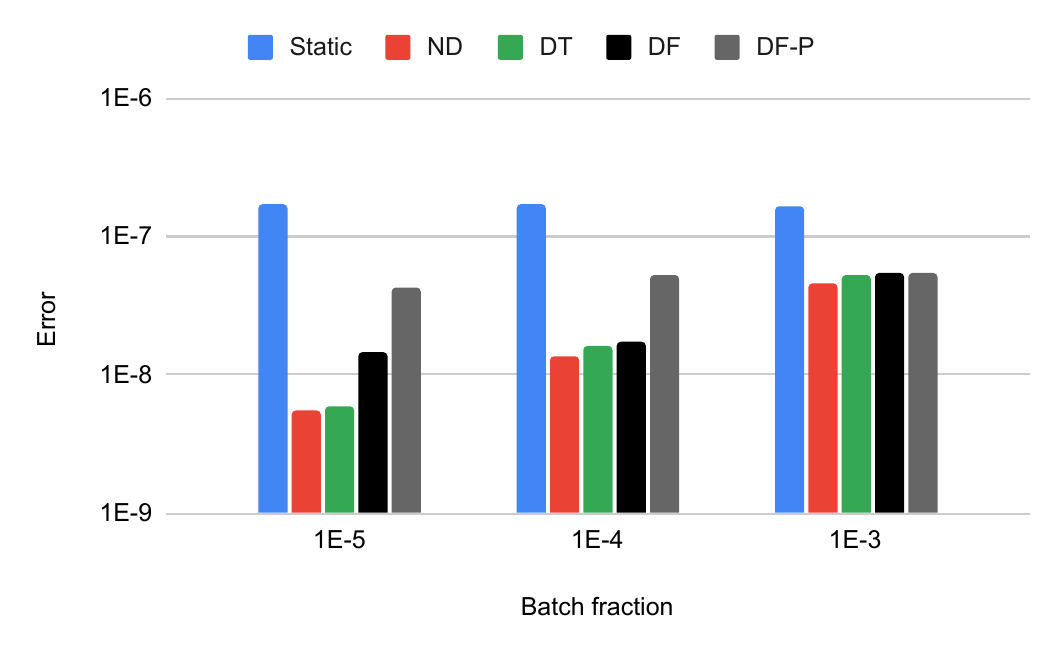}
  } \\[2ex]
  \includegraphics[width=0.48\linewidth]{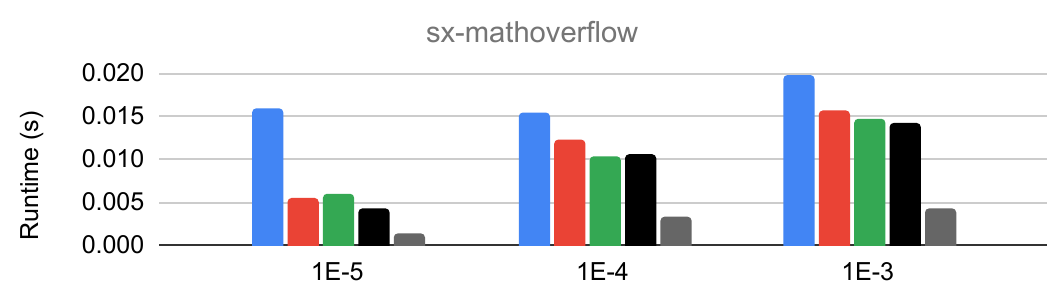}
  \includegraphics[width=0.48\linewidth]{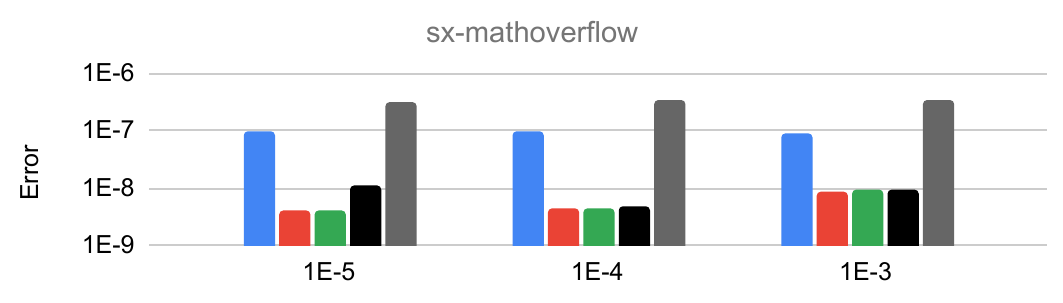}
  \includegraphics[width=0.48\linewidth]{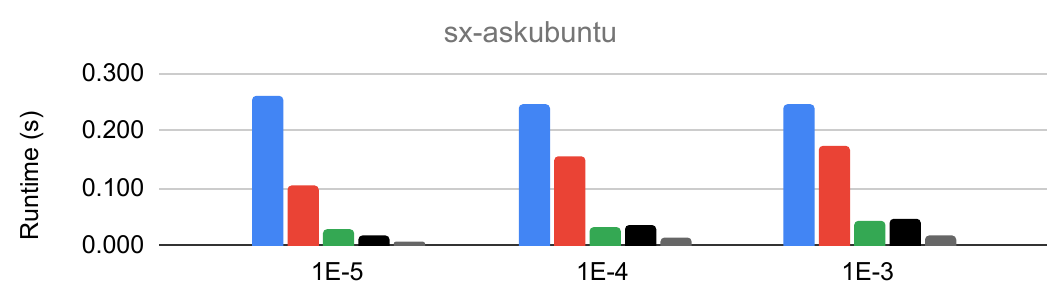}
  \includegraphics[width=0.48\linewidth]{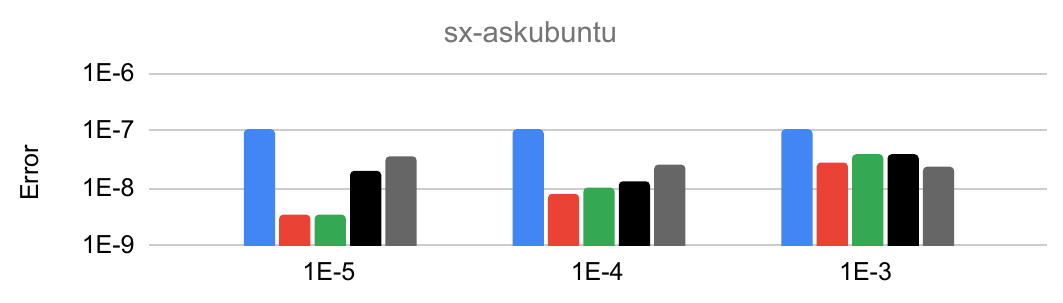}
  \includegraphics[width=0.48\linewidth]{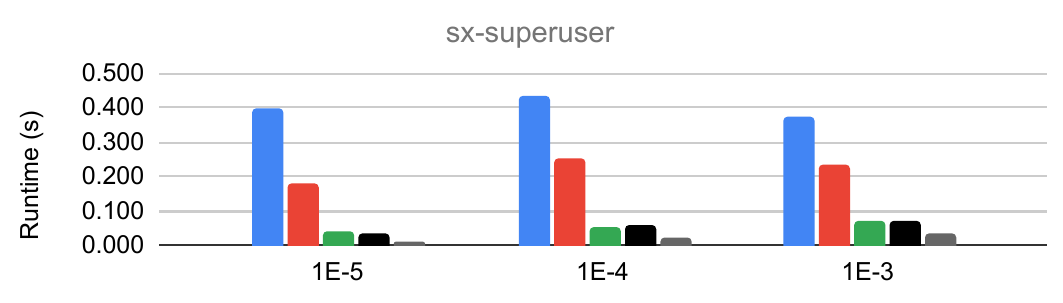}
  \includegraphics[width=0.48\linewidth]{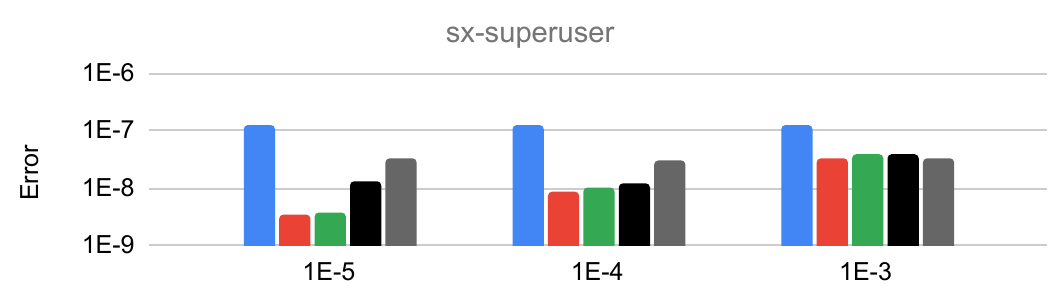}
  \includegraphics[width=0.48\linewidth]{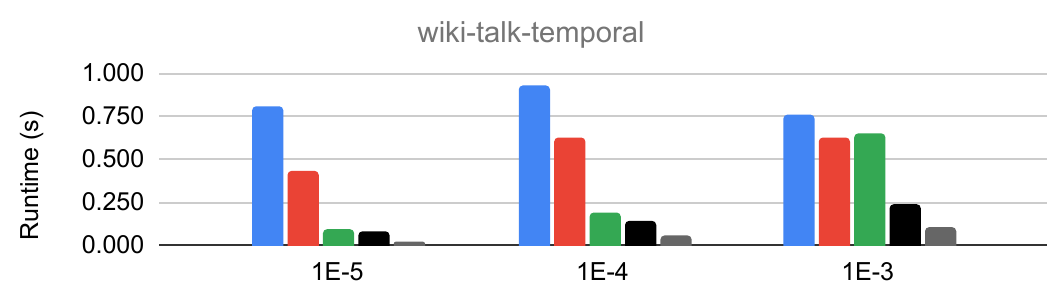}
  \includegraphics[width=0.48\linewidth]{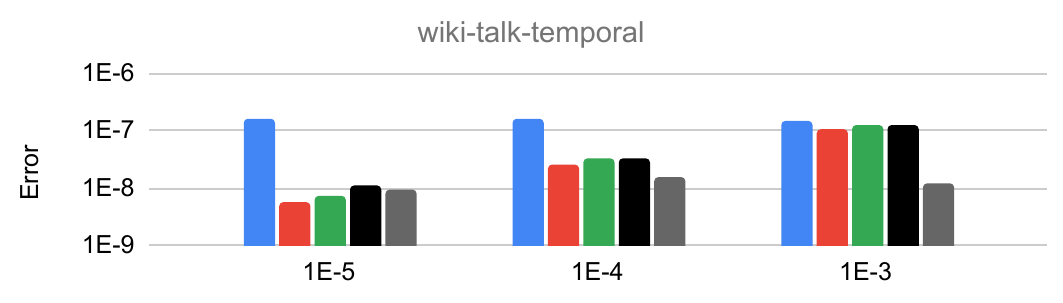}
  \subfigure[Runtime on each dynamic graph]{
    \label{fig:temporal-summary--runtime-graph}
    \includegraphics[width=0.48\linewidth]{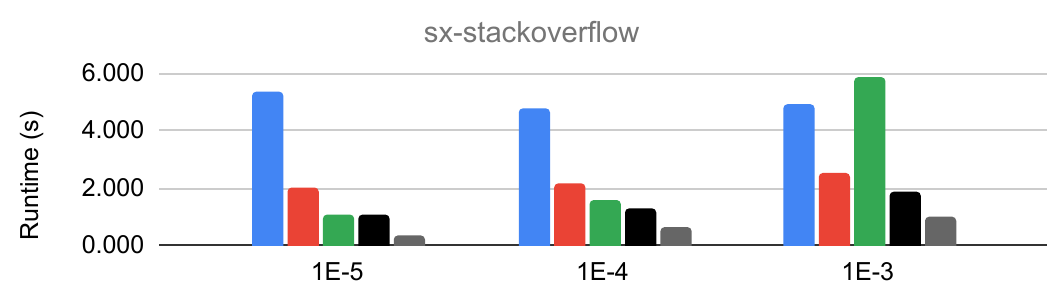}
  }
  \subfigure[Error in ranks obtained on each dynamic graph]{
    \label{fig:temporal-summary--error-graph}
    \includegraphics[width=0.48\linewidth]{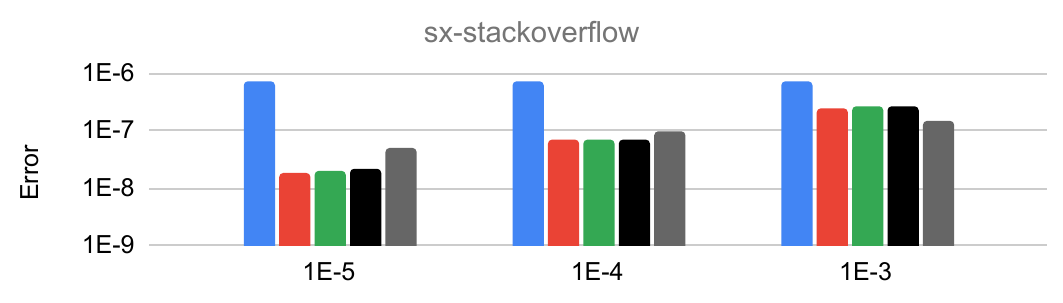}
  } \\[-2ex]
  \caption{Mean Runtime and Error in ranks obtained with \textit{Static}, \textit{Naive-dynamic (ND)}, \textit{Dynamic Traversal (DT)}, our improved \textit{Dynamic Frontier (DF)}, and our improved \textit{Dynamic Frontier with Pruning (DF-P)} PageRank on real-world dynamic graphs, with batch updates of size $10^{-5}|E_T|$ to $10^{-3}|E_T|$. Here, (a) and (b) show the overall runtime and error across all temporal graphs, while (c) and (d) show the runtime and rank error for each approach (relative to reference Static PageRank, see Section \ref{sec:measurement}). In (a), the speedup of each approach with respect to Static PageRank is labeled.}
  \label{fig:temporal-summary}
\end{figure*}

We now compare the performance of our improved Dynamic Frontier (DF) and Dynamic Frontier with Pruning (DF-P) PageRank algorithms with Static, Naive-dynamic (ND), and Dynamic Traversal (DT) PageRank on real-world dynamic graphs from Table \ref{tab:dataset}. This is done on batch updates of size $10^{-5}|E_T|$ to $10^{-3}|E_T|$ in multiples of $10$. For each batch size, we load $90\%$ of the graph initially and then load $B$ edges (where $B$ is the batch size) consecutively in $100$ batch updates. Self-loops are added to all vertices with each batch update. Figure \ref{fig:temporal-summary--runtime-overall} displays the overall runtime of each approach across all graphs for each batch size, while Figure \ref{fig:temporal-summary--error-overall} illustrates the overall rank error compared to a reference Static PageRank run (as described in Section \ref{sec:measurement}). Additionally, Figures \ref{fig:temporal-summary--runtime-graph} and \ref{fig:temporal-summary--error-graph} present the mean runtime and rank error of the approaches on each dynamic graph in the dataset. Finally, Figures \ref{fig:temporal-sx-mathoverflow}, \ref{fig:temporal-sx-askubuntu}, \ref{fig:temporal-sx-superuser}, \ref{fig:temporal-wiki-talk-temporal}, and \ref{fig:temporal-sx-stackoverflow} show the runtime and rank error of the approaches on each dynamic graph in Table \ref{tab:dataset}, upon each consecutive batch update.

Figure \ref{fig:temporal-summary--runtime-overall} shows that DF PageRank is, on average, $8.0\times$, $4.5\times$, and $3.2\times$ faster than Static PageRank for batch updates of size $10^{-5}|E_T|$, $10^{-4}|E_T|$, and $10^{-3}|E_T|$ respectively. Further, DF PageRank is, on average, $1.3\times$, $1.1\times$, and $1.5\times$ faster than DT PageRank, a widely used approach for updating PageRank on dynamic graphs, on the same batch updates. In contrast, DF-P PageRank is, on average, $26.2\times$, $11.9\times$, and $7.5\times$ faster than Static PageRank for batch updates of size $10^{-5}|E_T|$, $10^{-4}|E_T|$, and $10^{-3}|E_T|$ respectively. Furthermore, DF-P PageRank is, on average, $4.2\times$, $2.8\times$, and $3.6\times$ faster than DT PageRank on identical batch updates. This speedup is particularly higher on the \textit{sx-askubuntu} dynamic graph, with both DF and DF-P PageRank, as indicated by Figure \ref{fig:temporal-summary--runtime-graph}.

Regarding rank error, Figure \ref{fig:temporal-summary--error-overall} indicates that DF and DF-P PageRank have, on average, higher error than ND and DT PageRank but lower error than Static PageRank. This makes the ranks obtained with DF and DF-P PageRank acceptable. However, the error in ranks obtained with DF-P PageRank is consistently higher than that of Static PageRank on the \textit{sx-mathoverflow} dynamic graph (see Figure \ref{fig:temporal-summary--error-graph}), making DF PageRank the preferred approach on this particular graph. Therefore, DF-P PageRank can be the default choice for updating PageRank scores on dynamic graphs, but if higher error is observed (through intermediate empirical tests), switching to DF PageRank is recommended.

\subsubsection{Results on large graphs with random updates}

We also evaluate the performance of our improved Dynamic Frontier (DF) and Dynamic Frontier with Pruning (DF-P) PageRank algorithms alongside Static, Naive-dynamic (ND), and Dynamic Traversal (DT) PageRank on large (static) graphs from Table \ref{tab:dataset-large}, with randomly generated batch updates. This is done on batch updates of size $10^{-7}|E|$ to $0.1|E|$ (in multiples of $10$), comprising $80\%$ edge insertions and $20\%$ edge deletions in order to simulate realistic batch updates. Edge insertions are generated by selecting vertex pairs with equal probability, while edge deletions involve deleting each existing edge with a uniform probability. No new vertices are added to or removed from the graph, and self-loops are added to all vertices with each batch update. Figure \ref{fig:8020-runtime} illustrates the runtime of Static, ND, DT, DF, and DF-P PageRank, while Figure \ref{fig:8020-error} depicts the error in ranks obtained with each approach.

Figure \ref{fig:8020-runtime--mean} illustrates that for batch updates ranging from $10^{-7}|E|$ to $10^{-3}|E|$, comprising $80\%$ insertions and $20\%$ deletions, DF PageRank is, on average, $7.2\times$, $2.6\times$, and $4.0\times$ faster than Static, ND, and DT PageRank respectively. Additionally, DF-P PageRank is, on average, $9.6\times$, $3.9\times$, and $5.6\times$ faster than Static, ND, and DT PageRank respectively. This speedup is particularly higher on road networks and protein k-mer graphs, which have a low average degree (as depicted in Figure \ref{fig:8020-runtime--all}). It's worth noting that DT PageRank is slower than ND PageRank \cite{sahu2024incrementally} on large (static) graphs with uniformly random batch updates, as it ends up marking a large number of vertices as affected. This is due to updates being randomly scattered across the graph, leading to most of the graph being reachable from the updated regions.

Figure \ref{fig:8020-error--mean} indicates that DF-P PageRank generally exhibits higher error compared to ND, DT, and DF PageRank, but lower error than Static PageRank (up to a batch size of $10^{-2}|E|$). However, Figure \ref{fig:8020-error--all} highlights that the rank error with DF-P PageRank surpasses that of Static PageRank on web graphs. Consequently, DF PageRank is recommended as the preferred approach for web graphs with random batch updates.

\subsubsection{Comparison of vertices marked as affected}

\begin{figure}[!hbt]
  \centering
  \subfigure{
    \label{fig:measure-affected--batch}
    \includegraphics[width=0.98\linewidth]{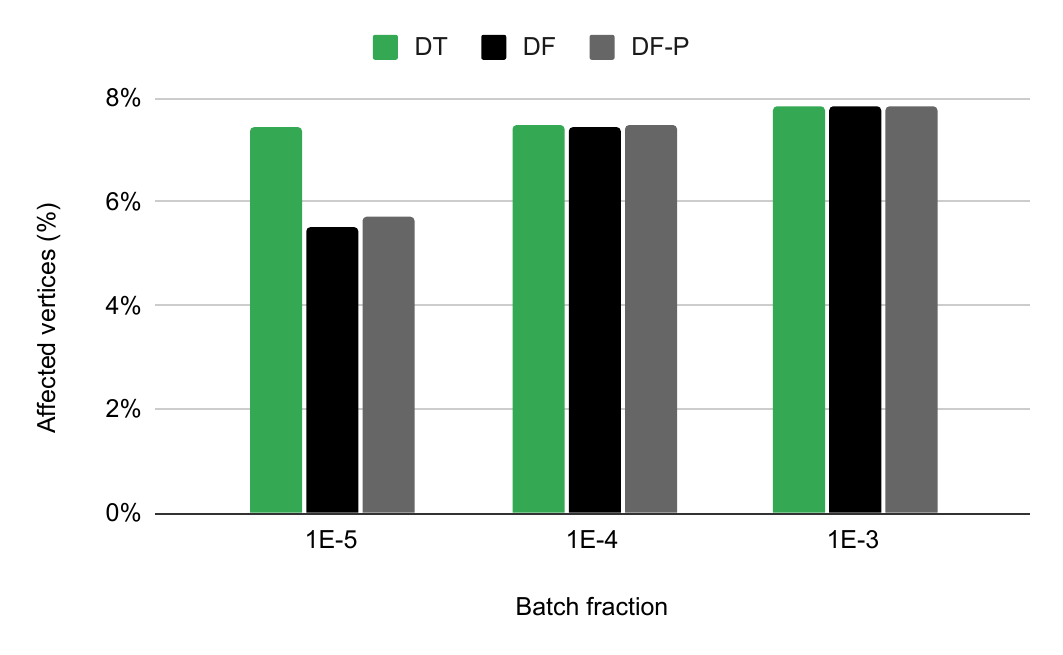}
  } \\[-2ex]
  \caption{Mean percentage of vertices marked as affected by \textit{Dynamic Traversal (DT)}, our improved \textit{Dynamic Frontier (DF)}, and \textit{Dynamic Frontier with Pruning (DF-P)} PageRank, on real-world graphs, with batch updates of size $10^{-5}|E_T|$ to $10^{-3}|E_T|$ (in multiples of $10$). DF and DF-P PageRank mark affected vertices incrementally --- thus, we count any vertex ever marked as affected.}
  \label{fig:measure-affected}
\end{figure}

Figure \ref{fig:measure-affected} displays the (mean) percentage of vertices marked as affected by Dynamic Traversal (DT), our improved Dynamic Frontier (DF), and Dynamic Frontier with Pruning (DF-P) PageRank on real-world dynamic graphs from Table \ref{tab:dataset}. This analysis is conducted on batch updates of size $10^{-5}|E_T|$ to $10^{-3}|E_T|$ in multiples of $10$ (see Section \ref{sec:batch-generation} for details). For DF and DF-P PageRank, affected vertices are marked incrementally --- therefore, we count all vertices that were ever flagged as affected.

As Figure \ref{fig:measure-affected} indicates, the proportion of vertices marked as affected by DF and DF-P PageRank is lower than DT PageRank for batch updates of size $10^{-5}|E_T|$, but comparable for larger batch updates. Therefore, the performance improvement with DF and DF-P PageRank is primarily attributed to the incremental marking of affected vertices. Additionally, it's worth noting that the percentage of vertices marked as affected is generally low across all approaches. This is likely because updates in real-world dynamic graphs tend to be concentrated in specific regions of the graph rather than being scattered throughout.

\begin{figure}[!hbt]
  \centering
  \subfigure{
    \label{fig:strong-scaling--speedup}
    \includegraphics[width=0.98\linewidth]{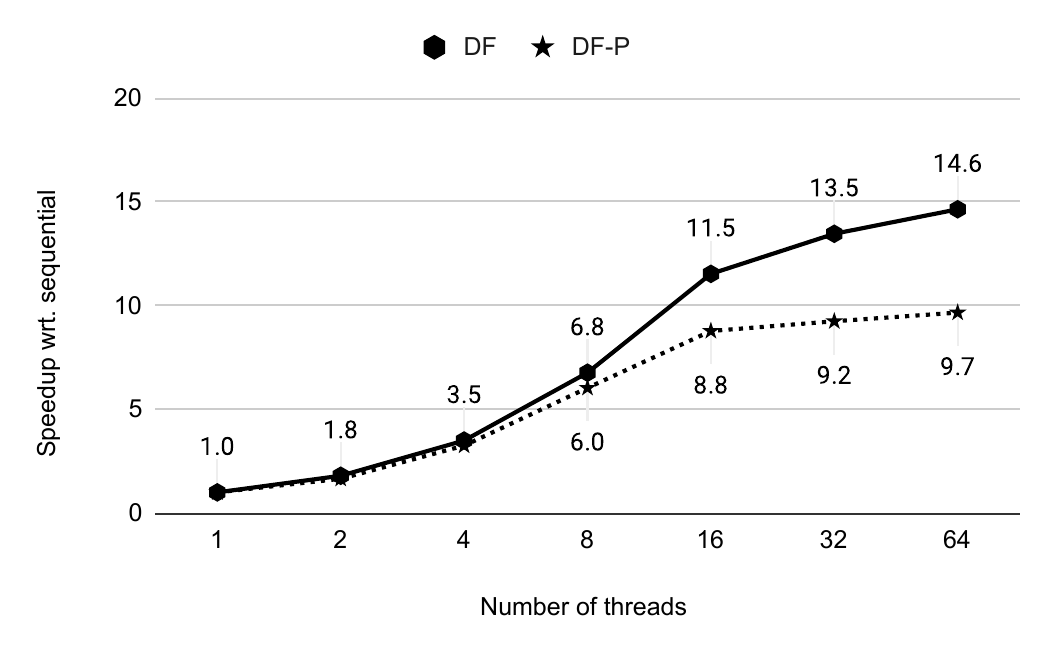}
  } \\[-2ex]
  \caption{Mean speedup of our improved \textit{Dynamic Frontier (DF)} and \textit{Dynamic Frontier with Pruning (DF-P)} PageRank with increasing number of threads (in multiples of $2$), on real-world dynamic graphs, with batch updates of size $10^{-4}|E_T|$.}
  \label{fig:strong-scaling}
\end{figure}

\subsection{Strong Scaling}

Finally, we examine the strong-scaling behavior of our improved Dynamic Frontier (DF) and Dynamic Frontier with Pruning (DF-P) PageRank algorithms on real-world dynamic graphs, with batch updates of a fixed size of $10^{-4}|E_T|$. The speedup of DF and DF-P PageRank is measured as the number of threads increases from $1$ to $64$ in multiples of $2$, relative to single-threaded execution. This process is repeated for each graph in the dataset (refer to Table \ref{tab:dataset}), and the results are averaged using geometric mean.

The results, depicted in Figure \ref{fig:strong-scaling}, indicate that with $16$ threads, DF PageRank achieves an average speedup of $11.5\times$ compared to single-threaded execution, showing a performance increase of $1.8\times$ for every doubling of threads. On the other hand, DF-P PageRank achieves an average speedup of $8.8\times$, suggesting a performance increase of $1.7\times$ for every doubling of threads. The speedup of DF-P PageRank is lower, likely due to the reduced work performed by the algorithm. At $32$ and $64$ threads, both DF and DF-P PageRank are affected by NUMA effects (the $64$-core processor used has $4$ NUMA domains), resulting in a speedup of only $13.5\times$ and $14.6\times$ for DF PageRank, and $9.2\times$ and $9.7\times$ for DF-P PageRank, respectively.

\section{Conclusion}
\label{sec:conclusion}
In conclusion, this study presents an efficient algorithm for updating PageRank on dynamic graphs. Given a batch update of edge insertions and deletions, our improved Dynamic Frontier (DF) and Dynamic Frontier with Pruning (DF-P) approaches identify an initial set of affected vertices and incrementally expand, and optionally contract/prune (with DF-P PageRank) this set across iterations. We observe that, expanding the frontier based on relative change in rank $\Delta r/r$ with a frontier tolerance $\tau_f$ of $10^{-6}$, and a corresponding prune tolerance $\tau_p$ of $10^{-6}$ (for DF-P PageRank) yields the best performance, while achieving lower error rates than Static PageRank.

On a server equipped with a 64-core AMD EPYC-7742 processor, DF PageRank demonstrates average speedups of $8.0\times$, $4.5\times$, and $3.2\times$ compared to Static PageRank when processing real-world dynamic graphs with batch updates of sizes $10^{-5}|E_T|$, $10^{-4}|E_T|$, and $10^{-3}|E_T|$, respectively. Additionally, it surpasses Dynamic Traversal (DT) PageRank, a commonly used method for updating PageRank on dynamic graphs, by $1.3\times$, $1.1\times$, and $1.5\times$ for the same batch updates. DF-P PageRank achieves even higher speedups, averaging $26.2\times$, $11.9\times$, and $7.5\times$ over Static PageRank, and $4.2\times$, $2.8\times$, and $3.6\times$ over DT PageRank for identical batch updates. For real-world dynamic graphs, we recommend DF-P PageRank, with a suggestion to switch to DF PageRank if higher error is observed.

For batch updates ranging from $10^{-7}|E|$ to $10^{-3}|E|$ with $80\%$ insertions and $20\%$ deletions on large static graphs, DF PageRank demonstrates average speedups of $7.2\times$, $2.6\times$, and $4.0\times$ compared to Static, ND, and DT PageRank respectively. Meanwhile, DF-P PageRank achieves average speedups of $9.6\times$, $3.9\times$, and $5.6\times$ over the same approaches. For large graphs with random updates, we recommend DF-P PageRank, except for web graphs, where we suggest selecting DF PageRank.

Using $16$ threads, DF PageRank exhibits an average speedup of $11.5\times$ compared to single-threaded execution, indicating a performance boost of $1.8\times$ for each doubling of threads. Conversely, DF-P PageRank achieves an average speedup of $8.8\times$, implying a performance increase of $1.7\times$ for each doubling of threads.

\begin{acks}
I would like to thank Prof. Kishore Kothapalli, Prof. Sathya Peri, and Prof. Hemalatha Eedi for their support.\ignore{Note that Britannia Industries Ltd., the owner of the 50-50 biscuit brand, did not sponsor our work.}
\end{acks}

\bibliographystyle{ACM-Reference-Format}
\bibliography{main}


\begin{thebibliography}{42}


\ifx \showCODEN    \undefined \def \showCODEN     #1{\unskip}     \fi
\ifx \showDOI      \undefined \def \showDOI       #1{#1}\fi
\ifx \showISBNx    \undefined \def \showISBNx     #1{\unskip}     \fi
\ifx \showISBNxiii \undefined \def \showISBNxiii  #1{\unskip}     \fi
\ifx \showISSN     \undefined \def \showISSN      #1{\unskip}     \fi
\ifx \showLCCN     \undefined \def \showLCCN      #1{\unskip}     \fi
\ifx \shownote     \undefined \def \shownote      #1{#1}          \fi
\ifx \showarticletitle \undefined \def \showarticletitle #1{#1}   \fi
\ifx \showURL      \undefined \def \showURL       {\relax}        \fi
\providecommand\bibfield[2]{#2}
\providecommand\bibinfo[2]{#2}
\providecommand\natexlab[1]{#1}
\providecommand\showeprint[2][]{arXiv:#2}

\bibitem[Agarwal et~al\mbox{.}(2012)]%
        {agarwal2012real}
\bibfield{author}{\bibinfo{person}{Manoj~K Agarwal}, \bibinfo{person}{Krithi Ramamritham}, {and} \bibinfo{person}{Manish Bhide}.} \bibinfo{year}{2012}\natexlab{}.
\newblock \showarticletitle{Real time discovery of dense clusters in highly dynamic graphs: identifying real world events in highly dynamic environments}.
\newblock \bibinfo{journal}{\emph{arXiv preprint arXiv:1207.0138}} (\bibinfo{year}{2012}).
\newblock


\bibitem[Allesina and Pascual(2009)]%
        {allesina2009googling}
\bibfield{author}{\bibinfo{person}{Stefano Allesina} {and} \bibinfo{person}{Mercedes Pascual}.} \bibinfo{year}{2009}\natexlab{}.
\newblock \showarticletitle{Googling food webs: can an eigenvector measure species' importance for coextinctions?}
\newblock \bibinfo{journal}{\emph{PLoS computational biology}} \bibinfo{volume}{5}, \bibinfo{number}{9} (\bibinfo{year}{2009}), \bibinfo{pages}{e1000494}.
\newblock


\bibitem[Andersen et~al\mbox{.}(2007)]%
        {rank-andersen07}
\bibfield{author}{\bibinfo{person}{R. Andersen}, \bibinfo{person}{F. Chung}, {and} \bibinfo{person}{K. Lang}.} \bibinfo{year}{2007}\natexlab{}.
\newblock \showarticletitle{Local partitioning for directed graphs using pagerank}. In \bibinfo{booktitle}{\emph{in Proc. WAW}}. \bibinfo{pages}{166--178}.
\newblock


\bibitem[Bahmani et~al\mbox{.}(2010)]%
        {bahmani2010fast}
\bibfield{author}{\bibinfo{person}{Bahman Bahmani}, \bibinfo{person}{Abdur Chowdhury}, {and} \bibinfo{person}{Ashish Goel}.} \bibinfo{year}{2010}\natexlab{}.
\newblock \showarticletitle{Fast incremental and personalized pagerank}.
\newblock \bibinfo{journal}{\emph{arXiv preprint arXiv:1006.2880}} (\bibinfo{year}{2010}).
\newblock


\bibitem[Bahmani et~al\mbox{.}(2012)]%
        {rank-bahmani12}
\bibfield{author}{\bibinfo{person}{Bahman Bahmani}, \bibinfo{person}{Ravi Kumar}, \bibinfo{person}{Mohammad Mahdian}, {and} \bibinfo{person}{Eli Upfal}.} \bibinfo{year}{2012}\natexlab{}.
\newblock \showarticletitle{Pagerank on an evolving graph}. In \bibinfo{booktitle}{\emph{Proceedings of the 18th ACM SIGKDD international conference on Knowledge discovery and data mining}}. \bibinfo{pages}{24--32}.
\newblock


\bibitem[B{\'a}nky et~al\mbox{.}(2013)]%
        {banky2013equal}
\bibfield{author}{\bibinfo{person}{D{\'a}niel B{\'a}nky}, \bibinfo{person}{G{\'a}bor Iv{\'a}n}, {and} \bibinfo{person}{Vince Grolmusz}.} \bibinfo{year}{2013}\natexlab{}.
\newblock \showarticletitle{Equal opportunity for low-degree network nodes: a PageRank-based method for protein target identification in metabolic graphs}.
\newblock \bibinfo{journal}{\emph{PLoS One}} \bibinfo{volume}{8}, \bibinfo{number}{1} (\bibinfo{year}{2013}), \bibinfo{pages}{e54204}.
\newblock


\bibitem[Barros et~al\mbox{.}(2021)]%
        {barros2021survey}
\bibfield{author}{\bibinfo{person}{Claudio~DT Barros}, \bibinfo{person}{Matheus~RF Mendon{\c{c}}a}, \bibinfo{person}{Alex~B Vieira}, {and} \bibinfo{person}{Artur Ziviani}.} \bibinfo{year}{2021}\natexlab{}.
\newblock \showarticletitle{A survey on embedding dynamic graphs}.
\newblock \bibinfo{journal}{\emph{ACM Computing Surveys (CSUR)}} \bibinfo{volume}{55}, \bibinfo{number}{1} (\bibinfo{year}{2021}), \bibinfo{pages}{1--37}.
\newblock


\bibitem[Berberich et~al\mbox{.}(2007)]%
        {rank-berberich07}
\bibfield{author}{\bibinfo{person}{Klaus Berberich}, \bibinfo{person}{Srikanta Bedathur}, \bibinfo{person}{Gerhard Weikum}, {and} \bibinfo{person}{Michalis Vazirgiannis}.} \bibinfo{year}{2007}\natexlab{}.
\newblock \showarticletitle{Comparing apples and oranges: normalized pagerank for evolving graphs}. In \bibinfo{booktitle}{\emph{Proceedings of the 16th international conference on world wide web}}. \bibinfo{pages}{1145--1146}.
\newblock


\bibitem[Chen et~al\mbox{.}(2004)]%
        {chen2004local}
\bibfield{author}{\bibinfo{person}{Yen-Yu Chen}, \bibinfo{person}{Qingqing Gan}, {and} \bibinfo{person}{Torsten Suel}.} \bibinfo{year}{2004}\natexlab{}.
\newblock \showarticletitle{Local methods for estimating pagerank values}. In \bibinfo{booktitle}{\emph{Proceedings of the thirteenth ACM international conference on Information and knowledge management}}. \bibinfo{pages}{381--389}.
\newblock


\bibitem[Chepelianskii(2010)]%
        {chepelianskii2010towards}
\bibfield{author}{\bibinfo{person}{Alexei~D Chepelianskii}.} \bibinfo{year}{2010}\natexlab{}.
\newblock \showarticletitle{Towards physical laws for software architecture}.
\newblock \bibinfo{journal}{\emph{arXiv preprint arXiv:1003.5455}} (\bibinfo{year}{2010}).
\newblock


\bibitem[Chien et~al\mbox{.}(2001)]%
        {rank-chien01}
\bibfield{author}{\bibinfo{person}{S. Chien}, \bibinfo{person}{C. Dwork}, \bibinfo{person}{R. Kumar}, {and} \bibinfo{person}{D. Sivakumar}.} \bibinfo{year}{2001}\natexlab{}.
\newblock \bibinfo{title}{Towards Exploiting Link Evolution}.
\newblock
\newblock


\bibitem[Desikan et~al\mbox{.}(2005)]%
        {rank-desikan05}
\bibfield{author}{\bibinfo{person}{P. Desikan}, \bibinfo{person}{N. Pathak}, \bibinfo{person}{J. Srivastava}, {and} \bibinfo{person}{V. Kumar}.} \bibinfo{year}{2005}\natexlab{}.
\newblock \showarticletitle{{Incremental Page Rank Computation on Evolving Graphs}}. In \bibinfo{booktitle}{\emph{Special Interest Tracks and Posters of the 14th International Conference on World Wide Web}} (Chiba, Japan) \emph{(\bibinfo{series}{WWW '05})}. \bibinfo{publisher}{Association for Computing Machinery}, \bibinfo{address}{New York, NY, USA}, \bibinfo{pages}{1094--1095}.
\newblock
\showISBNx{1595930515}
\urldef\tempurl%
\url{https://doi.org/10.1145/1062745.1062885}
\showURL{%
\tempurl}


\bibitem[Dhulipala et~al\mbox{.}(2019)]%
        {graph-dhulipala19}
\bibfield{author}{\bibinfo{person}{L. Dhulipala}, \bibinfo{person}{G.E. Blelloch}, {and} \bibinfo{person}{J. Shun}.} \bibinfo{year}{2019}\natexlab{}.
\newblock \showarticletitle{Low-latency graph streaming using compressed purely-functional trees}. In \bibinfo{booktitle}{\emph{ACM SIGPLAN PLDI}}. \bibinfo{pages}{918--934}.
\newblock


\bibitem[Dubey and Khare(2022)]%
        {rank-dubey22}
\bibfield{author}{\bibinfo{person}{H. Dubey} {and} \bibinfo{person}{N. Khare}.} \bibinfo{year}{2022}\natexlab{}.
\newblock \showarticletitle{Fast parallel computation of PageRank scores with improved convergence time}.
\newblock \bibinfo{journal}{\emph{IJDMMM}} \bibinfo{volume}{14}, \bibinfo{number}{1} (\bibinfo{year}{2022}), \bibinfo{pages}{63--88}.
\newblock


\bibitem[Eppstein et~al\mbox{.}(1997)]%
        {graph-eppstein97}
\bibfield{author}{\bibinfo{person}{D. Eppstein}, \bibinfo{person}{Z. Galil}, \bibinfo{person}{G. Italiano}, {and} \bibinfo{person}{A. Nissenzweig}.} \bibinfo{year}{1997}\natexlab{}.
\newblock \showarticletitle{{Sparsification — A technique for speeding up dynamic graph algorithms}}.
\newblock \bibinfo{journal}{\emph{J. ACM}} \bibinfo{volume}{44}, \bibinfo{number}{5} (\bibinfo{date}{September} \bibinfo{year}{1997}), \bibinfo{pages}{669--696}.
\newblock
\urldef\tempurl%
\url{http://doi.acm.org/10.1145/265910.265914}
\showURL{%
\tempurl}


\bibitem[Fender et~al\mbox{.}({[n.\,d.]})]%
        {rank-nvgraph}
\bibfield{author}{\bibinfo{person}{A. Fender}, \bibinfo{person}{N. Thejaswi}, {and} \bibinfo{person}{B. Rees}.} \bibinfo{year}{[n.\,d.]}\natexlab{}.
\newblock \bibinfo{title}{{rapidsai/nvgraph}}.
\newblock
\newblock
\urldef\tempurl%
\url{https://github.com/rapidsai/nvgraph/blob/main/cpp/src/pagerank.cu#L149}
\showURL{%
\tempurl}


\bibitem[Garg and Kothapalli(2016)]%
        {rank-garg16}
\bibfield{author}{\bibinfo{person}{P. Garg} {and} \bibinfo{person}{K. Kothapalli}.} \bibinfo{year}{2016}\natexlab{}.
\newblock \showarticletitle{{STIC-D: Algorithmic Techniques For Efficient Parallel Pagerank Computation on Real-World Graphs}}. In \bibinfo{booktitle}{\emph{Proceedings of the 17th International Conference on Distributed Computing and Networking - ICDCN ’16}}. \bibinfo{publisher}{ACM Press}, \bibinfo{pages}{1---10}.
\newblock
\showISBNx{9781450340328}


\bibitem[Giri et~al\mbox{.}(2020)]%
        {rank-giri20}
\bibfield{author}{\bibinfo{person}{H. Giri}, \bibinfo{person}{M. Haque}, {and} \bibinfo{person}{D. Banerjee}.} \bibinfo{year}{2020}\natexlab{}.
\newblock \showarticletitle{{HyPR: Hybrid Page Ranking on Evolving Graphs}}. In \bibinfo{booktitle}{\emph{Proc. IEEE 27th International Conference on High Performance Computing, Data, and Analytics (HiPC)}}. \bibinfo{pages}{62--71}.
\newblock


\bibitem[Guoqiang et~al\mbox{.}(2020)]%
        {rank-guoqiang20}
\bibfield{author}{\bibinfo{person}{M. Guoqiang}, \bibinfo{person}{H. Rui}, \bibinfo{person}{W. Jiangwei}, \bibinfo{person}{K. Hongwei}, {and} \bibinfo{person}{L. Rengang}.} \bibinfo{year}{2020}\natexlab{}.
\newblock \showarticletitle{{A FPGA based intra-parallel architecture for PageRank graph processing}}. In \bibinfo{booktitle}{\emph{IEEE International Conference on Edge Computing (EDGE)}}. \bibinfo{publisher}{IEEE}, \bibinfo{pages}{31--38}.
\newblock
\showISBNx{978-1-7281-8254-4}


\bibitem[Kim and Choi(2015)]%
        {kim2015incremental}
\bibfield{author}{\bibinfo{person}{Kyung~Soo Kim} {and} \bibinfo{person}{Yong~Suk Choi}.} \bibinfo{year}{2015}\natexlab{}.
\newblock \showarticletitle{Incremental iteration method for fast pagerank computation}. In \bibinfo{booktitle}{\emph{Proceedings of the 9th International Conference on Ubiquitous Information Management and Communication}}. \bibinfo{pages}{1--5}.
\newblock


\bibitem[Kim et~al\mbox{.}(2015)]%
        {traffic-kim15}
\bibfield{author}{\bibinfo{person}{Y. Kim}, \bibinfo{person}{H. Kim}, \bibinfo{person}{C. Shin}, \bibinfo{person}{K. Lee}, \bibinfo{person}{C. Choi}, {and} \bibinfo{person}{W. Cho}.} \bibinfo{year}{2015}\natexlab{}.
\newblock \showarticletitle{{Analysis on the transportation point in cheongju city using pagerank algorithm}}. In \bibinfo{booktitle}{\emph{Proceedings of the International Conference on Big Data Applications and Services - BigDAS ’15}}, \bibfield{editor}{\bibinfo{person}{C.~Leung} {and} \bibinfo{person}{A.~Nasridinov}} (Eds.). \bibinfo{publisher}{ACM Press}, \bibinfo{address}{New York, New York, USA}, \bibinfo{pages}{165--169}.
\newblock
\showISBNx{9781450338462}


\bibitem[Kolda and Procopio(2009)]%
        {kolda2009generalized}
\bibfield{author}{\bibinfo{person}{Tamara~G Kolda} {and} \bibinfo{person}{Michael~J Procopio}.} \bibinfo{year}{2009}\natexlab{}.
\newblock \showarticletitle{Generalized badrank with graduated trust}.
\newblock \bibinfo{journal}{\emph{Sandia National Laboratories, California}} (\bibinfo{year}{2009}).
\newblock


\bibitem[Kolodziej et~al\mbox{.}(2019)]%
        {suite19}
\bibfield{author}{\bibinfo{person}{S. Kolodziej}, \bibinfo{person}{M. Aznaveh}, \bibinfo{person}{M. Bullock}, \bibinfo{person}{J. David}, \bibinfo{person}{T. Davis}, \bibinfo{person}{M. Henderson}, \bibinfo{person}{Y. Hu}, {and} \bibinfo{person}{R. Sandstrom}.} \bibinfo{year}{2019}\natexlab{}.
\newblock \showarticletitle{{The SuiteSparse matrix collection website interface}}.
\newblock \bibinfo{journal}{\emph{The Journal of Open Source Software}} \bibinfo{volume}{4}, \bibinfo{number}{35} (\bibinfo{date}{Mar} \bibinfo{year}{2019}), \bibinfo{pages}{1244}.
\newblock


\bibitem[Langville and Meyer(2006)]%
        {rank-langville06}
\bibfield{author}{\bibinfo{person}{A.N. Langville} {and} \bibinfo{person}{C.D. Meyer}.} \bibinfo{year}{2006}\natexlab{}.
\newblock \showarticletitle{A reordering for the PageRank problem}.
\newblock \bibinfo{journal}{\emph{SIAM SISC}} \bibinfo{volume}{27}, \bibinfo{number}{6} (\bibinfo{year}{2006}), \bibinfo{pages}{2112--2120}.
\newblock


\bibitem[Leskovec and Krevl(2014)]%
        {snapnets}
\bibfield{author}{\bibinfo{person}{Jure Leskovec} {and} \bibinfo{person}{Andrej Krevl}.} \bibinfo{year}{2014}\natexlab{}.
\newblock \bibinfo{title}{{SNAP Datasets}: {Stanford} Large Network Dataset Collection}.
\newblock \bibinfo{howpublished}{\url{http://snap.stanford.edu/data}}.
\newblock


\bibitem[Li et~al\mbox{.}(2021)]%
        {rank-li21}
\bibfield{author}{\bibinfo{person}{L. Li}, \bibinfo{person}{Y. Chen}, \bibinfo{person}{Z. Zirnheld}, \bibinfo{person}{P. Li}, {and} \bibinfo{person}{C. Hao}.} \bibinfo{year}{2021}\natexlab{}.
\newblock \showarticletitle{{MELOPPR: Software/Hardware Co-design for Memory-efficient Low-latency Personalized PageRank}}.
\newblock  (\bibinfo{year}{2021}).
\newblock


\bibitem[{NVIDIA Corporation}(2019)]%
        {nvgraph}
\bibfield{author}{\bibinfo{person}{{NVIDIA Corporation}}.} \bibinfo{year}{2019}\natexlab{}.
\newblock \bibinfo{title}{{nvGRAPH Library User's Guide}}.
\newblock
\newblock
\urldef\tempurl%
\url{https://docs.nvidia.com/cuda/archive/10.1/pdf/nvGRAPH_Library.pdf}
\showURL{%
\tempurl}


\bibitem[Ohsaka et~al\mbox{.}(2015)]%
        {ohsaka2015efficient}
\bibfield{author}{\bibinfo{person}{Naoto Ohsaka}, \bibinfo{person}{Takanori Maehara}, {and} \bibinfo{person}{Ken-ichi Kawarabayashi}.} \bibinfo{year}{2015}\natexlab{}.
\newblock \showarticletitle{Efficient pagerank tracking in evolving networks}. In \bibinfo{booktitle}{\emph{Proceedings of the 21th ACM SIGKDD international conference on knowledge discovery and data mining}}. \bibinfo{pages}{875--884}.
\newblock


\bibitem[Page et~al\mbox{.}(1999)]%
        {rank-page99}
\bibfield{author}{\bibinfo{person}{L. Page}, \bibinfo{person}{S. Brin}, \bibinfo{person}{R. Motwani}, {and} \bibinfo{person}{T. Winograd}.} \bibinfo{year}{1999}\natexlab{}.
\newblock \bibinfo{booktitle}{\emph{{The PageRank citation ranking: Bringing order to the web.}}}
\newblock \bibinfo{type}{{T}echnical {R}eport}. \bibinfo{institution}{Stanford InfoLab}.
\newblock


\bibitem[Pashikanti and Kundu(2022)]%
        {rank-pashikanti22}
\bibfield{author}{\bibinfo{person}{R.P. Pashikanti} {and} \bibinfo{person}{S. Kundu}.} \bibinfo{year}{2022}\natexlab{}.
\newblock \showarticletitle{FPPR: fast pessimistic (dynamic) PageRank to update PageRank in evolving directed graphs on network changes}.
\newblock \bibinfo{journal}{\emph{SNAM}} \bibinfo{volume}{12}, \bibinfo{number}{1} (\bibinfo{year}{2022}), \bibinfo{pages}{141}.
\newblock


\bibitem[Plimpton and Devine(2011)]%
        {rank-plimpton11}
\bibfield{author}{\bibinfo{person}{S.J. Plimpton} {and} \bibinfo{person}{K.D. Devine}.} \bibinfo{year}{2011}\natexlab{}.
\newblock \showarticletitle{MapReduce in MPI for large-scale graph algorithms}.
\newblock \bibinfo{journal}{\emph{Parallel Comput.}} \bibinfo{volume}{37}, \bibinfo{number}{9} (\bibinfo{year}{2011}), \bibinfo{pages}{610--632}.
\newblock


\bibitem[Ramalingam(1996)]%
        {incr-ramalingam96}
\bibfield{author}{\bibinfo{person}{G. Ramalingam}.} \bibinfo{year}{1996}\natexlab{}.
\newblock \showarticletitle{{Bounded Incremental Computation}}.
\newblock \bibinfo{journal}{\emph{Lecture Notes in Computer Science}}  \bibinfo{volume}{1089} (\bibinfo{year}{1996}), \bibinfo{pages}{101--129}.
\newblock
\showISBNx{3-540-61320-X}


\bibitem[Sadi et~al\mbox{.}(2018)]%
        {rank-sadi18}
\bibfield{author}{\bibinfo{person}{F. Sadi}, \bibinfo{person}{J. Sweeney}, \bibinfo{person}{S. McMillan}, \bibinfo{person}{T. Low}, \bibinfo{person}{J. Hoe}, \bibinfo{person}{L. Pileggi}, {and} \bibinfo{person}{F. Franchetti}.} \bibinfo{year}{2018}\natexlab{}.
\newblock \showarticletitle{{PageRank Acceleration for Large Graphs with Scalable Hardware and Two-Step SpMV}}. In \bibinfo{booktitle}{\emph{IEEE High Performance extreme Computing Conference (HPEC)}}. \bibinfo{publisher}{IEEE}, \bibinfo{pages}{1--7}.
\newblock
\showISBNx{978-1-5386-5989-2}


\bibitem[Sahu(2024)]%
        {sahu2024incrementally}
\bibfield{author}{\bibinfo{person}{Subhajit Sahu}.} \bibinfo{year}{2024}\natexlab{}.
\newblock \showarticletitle{An Incrementally Expanding Approach for Updating PageRank on Dynamic Graphs}.
\newblock \bibinfo{journal}{\emph{arXiv preprint arXiv:2401.03256}} (\bibinfo{year}{2024}).
\newblock


\bibitem[Sahu et~al\mbox{.}(2022)]%
        {sahu2022dynamic}
\bibfield{author}{\bibinfo{person}{Subhajit Sahu}, \bibinfo{person}{Kishore Kothapalli}, {and} \bibinfo{person}{Dip~Sankar Banerjee}.} \bibinfo{year}{2022}\natexlab{}.
\newblock \showarticletitle{Dynamic Batch Parallel Algorithms for Updating PageRank}. In \bibinfo{booktitle}{\emph{2022 IEEE International Parallel and Distributed Processing Symposium Workshops (IPDPSW)}}. IEEE, \bibinfo{pages}{1129--1138}.
\newblock


\bibitem[Sarma et~al\mbox{.}(2013)]%
        {rank-sarma13}
\bibfield{author}{\bibinfo{person}{A. Sarma}, \bibinfo{person}{A. Molla}, \bibinfo{person}{G. Pandurangan}, {and} \bibinfo{person}{E. Upfal}.} \bibinfo{year}{2013}\natexlab{}.
\newblock \showarticletitle{{Fast Distributed PageRank Computation}}. In \bibinfo{booktitle}{\emph{Distributed Computing and Networking}}. \bibinfo{publisher}{Springer Berlin Heidelberg}, \bibinfo{address}{Berlin, Heidelberg}, \bibinfo{pages}{11--26}.
\newblock


\bibitem[Senanayake et~al\mbox{.}(2015)]%
        {rank-senanayake15}
\bibfield{author}{\bibinfo{person}{U. Senanayake}, \bibinfo{person}{M. Piraveenan}, {and} \bibinfo{person}{A. Zomaya}.} \bibinfo{year}{2015}\natexlab{}.
\newblock \showarticletitle{{The pagerank-index: Going beyond citation counts in quantifying scientific impact of researchers}}.
\newblock \bibinfo{journal}{\emph{PloS one}} \bibinfo{volume}{10}, \bibinfo{number}{8} (\bibinfo{year}{2015}), \bibinfo{pages}{e0134794}.
\newblock


\bibitem[Verstraaten et~al\mbox{.}(2015)]%
        {verstraaten2015quantifying}
\bibfield{author}{\bibinfo{person}{Merijn Verstraaten}, \bibinfo{person}{Ana~Lucia Varbanescu}, {and} \bibinfo{person}{Cees de Laat}.} \bibinfo{year}{2015}\natexlab{}.
\newblock \showarticletitle{Quantifying the performance impact of graph structure on neighbour iteration strategies for pagerank}. In \bibinfo{booktitle}{\emph{Euro-Par 2015: Parallel Processing Workshops: Euro-Par 2015 International Workshops, Vienna, Austria, August 24-25, 2015, Revised Selected Papers 21}}. Springer, \bibinfo{pages}{528--540}.
\newblock


\bibitem[Zhan et~al\mbox{.}(2019)]%
        {zhan2019fast}
\bibfield{author}{\bibinfo{person}{Zexing Zhan}, \bibinfo{person}{Ruimin Hu}, \bibinfo{person}{Xiyue Gao}, {and} \bibinfo{person}{Nian Huai}.} \bibinfo{year}{2019}\natexlab{}.
\newblock \showarticletitle{Fast incremental pagerank on dynamic networks}. In \bibinfo{booktitle}{\emph{International Conference on Web Engineering}}. Springer, \bibinfo{pages}{154--168}.
\newblock


\bibitem[Zhang and Yuan(2018)]%
        {urban-zhang18}
\bibfield{author}{\bibinfo{person}{Q. Zhang} {and} \bibinfo{person}{T. Yuan}.} \bibinfo{year}{2018}\natexlab{}.
\newblock \showarticletitle{{Analysis of China’s Urban Network Structure from the Perspective of “Streaming”}}. In \bibinfo{booktitle}{\emph{26th International Conference on Geoinformatics}}. \bibinfo{publisher}{IEEE}, \bibinfo{pages}{1--7}.
\newblock
\showISBNx{978-1-5386-7619-6}


\bibitem[Zhang(2017)]%
        {rank-zhang17}
\bibfield{author}{\bibinfo{person}{T. Zhang}.} \bibinfo{year}{2017}\natexlab{}.
\newblock \showarticletitle{Efficient incremental pagerank of evolving graphs on GPU}. In \bibinfo{booktitle}{\emph{IEEE ICCSEC}}. \bibinfo{pages}{1232--1236}.
\newblock


\bibitem[Zuo et~al\mbox{.}(2012)]%
        {zuo2012network}
\bibfield{author}{\bibinfo{person}{Xi-Nian Zuo}, \bibinfo{person}{Ross Ehmke}, \bibinfo{person}{Maarten Mennes}, \bibinfo{person}{Davide Imperati}, \bibinfo{person}{F~Xavier Castellanos}, \bibinfo{person}{Olaf Sporns}, {and} \bibinfo{person}{Michael~P Milham}.} \bibinfo{year}{2012}\natexlab{}.
\newblock \showarticletitle{Network centrality in the human functional connectome}.
\newblock \bibinfo{journal}{\emph{Cerebral cortex}} \bibinfo{volume}{22}, \bibinfo{number}{8} (\bibinfo{year}{2012}), \bibinfo{pages}{1862--1875}.
\newblock


\end{thebibliography}

\clearpage
\appendix
\begin{figure*}[!hbt]
  \centering
  \subfigure[Runtime on consecutive batch updates of size $10^{-5}|E_T|$]{
    \label{fig:temporal-sx-mathoverflow--runtime5}
    \includegraphics[width=0.48\linewidth]{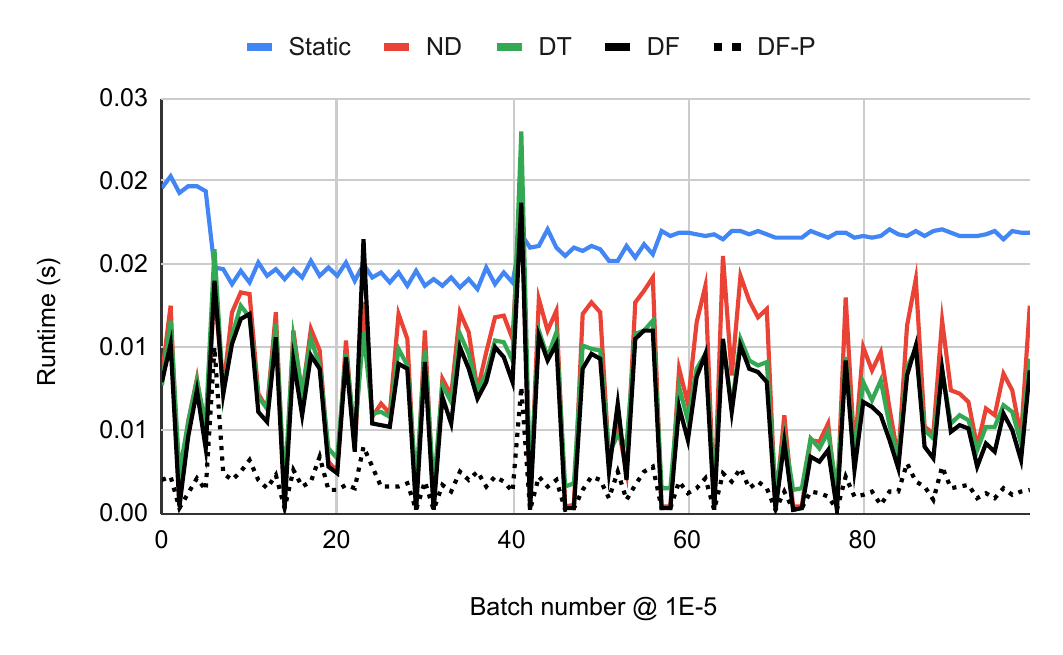}
  }
  \subfigure[Error in ranks obtained on consecutive batch updates of size $10^{-5}|E_T|$]{
    \label{fig:temporal-sx-mathoverflow--error5}
    \includegraphics[width=0.48\linewidth]{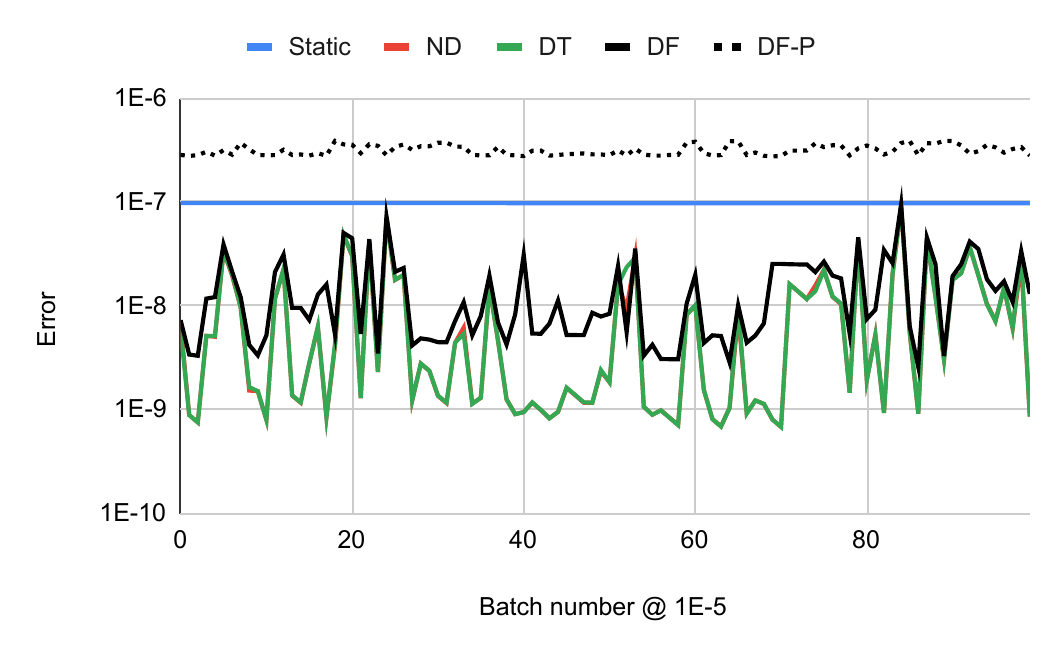}
  } \\[2ex]
  \subfigure[Runtime on consecutive batch updates of size $10^{-4}|E_T|$]{
    \label{fig:temporal-sx-mathoverflow--runtime4}
    \includegraphics[width=0.48\linewidth]{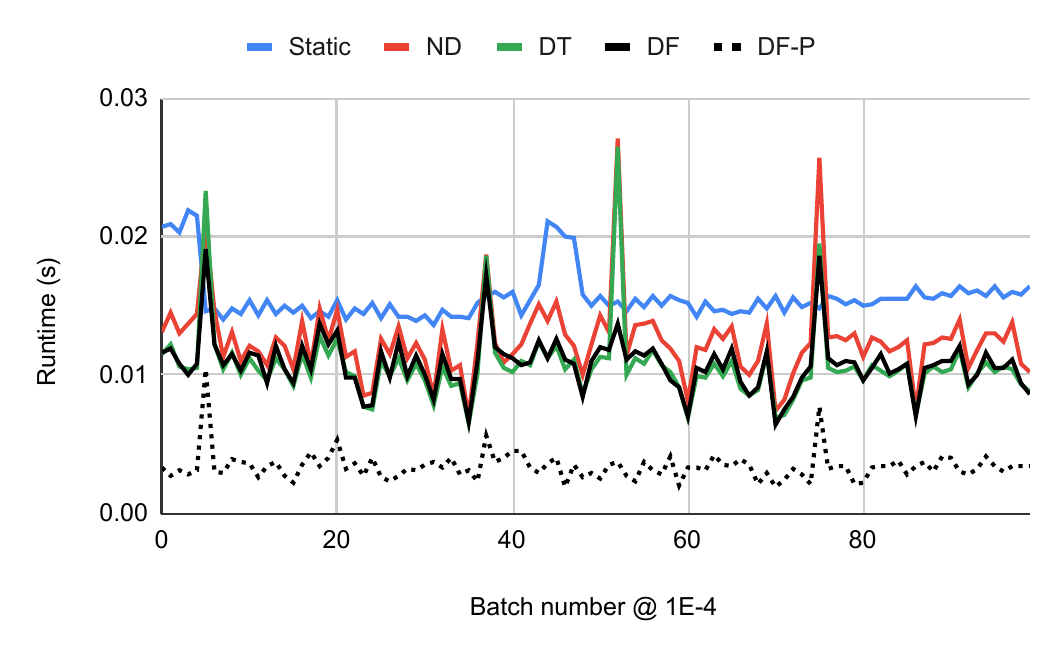}
  }
  \subfigure[Error in ranks obtained on consecutive batch updates of size $10^{-4}|E_T|$]{
    \label{fig:temporal-sx-mathoverflow--error4}
    \includegraphics[width=0.48\linewidth]{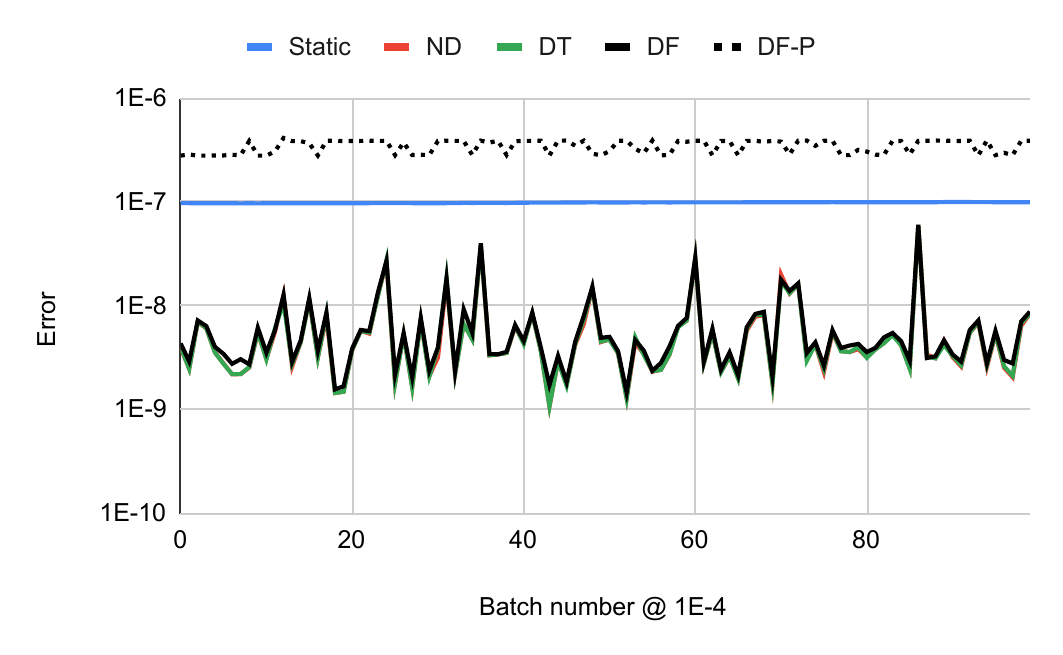}
  } \\[2ex]
  \subfigure[Runtime on consecutive batch updates of size $10^{-3}|E_T|$]{
    \label{fig:temporal-sx-mathoverflow--runtime3}
    \includegraphics[width=0.48\linewidth]{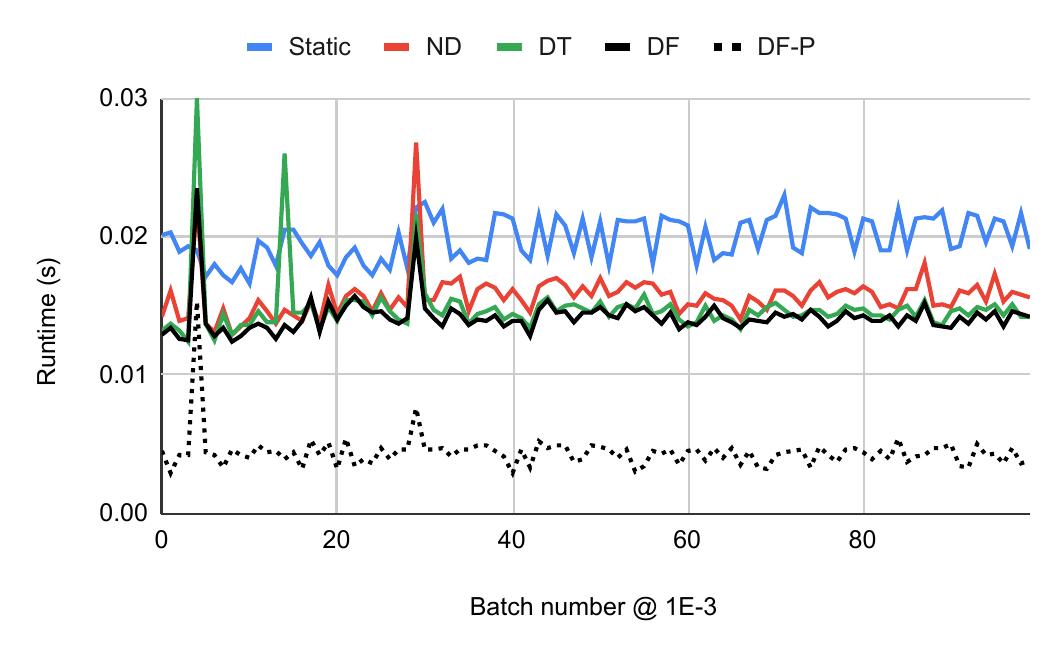}
  }
  \subfigure[Error in ranks obtained on consecutive batch updates of size $10^{-3}|E_T|$]{
    \label{fig:temporal-sx-mathoverflow--error3}
    \includegraphics[width=0.48\linewidth]{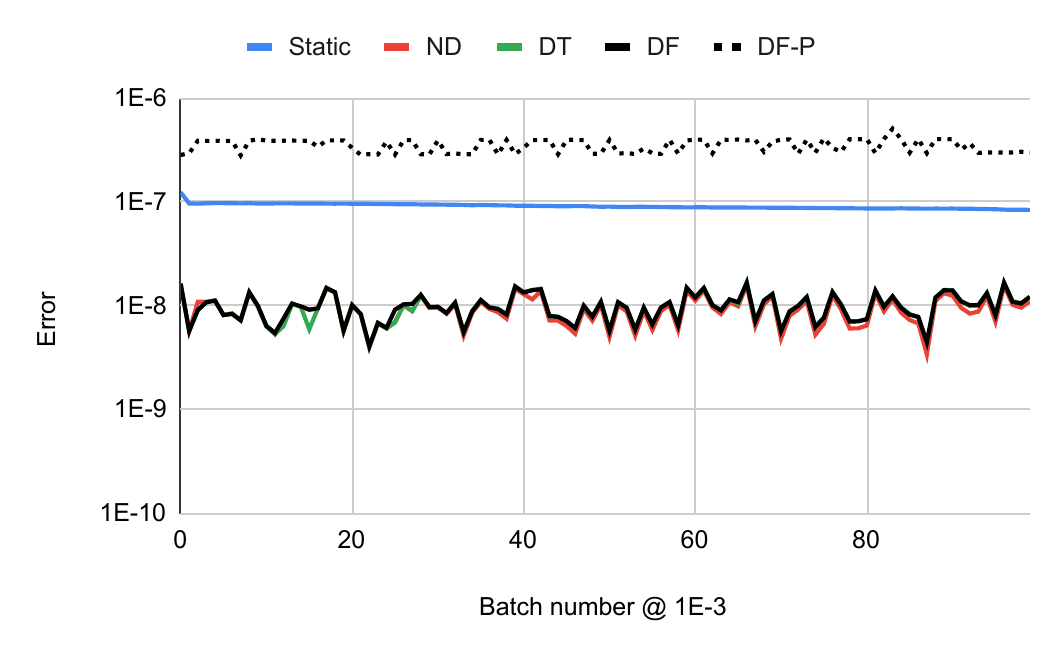}
  } \\[-2ex]
  \caption{Runtime and Error in ranks obtained with \textit{Static}, \textit{Naive-dynamic (ND)}, \textit{Dynamic Traversal (DT)}, our improved \textit{Dynamic Frontier (DF)}, and our improved \textit{Dynamic Frontier with Pruning (DF-P)} PageRank on the \textit{sx-mathoverflow} dynamic graph. The size of batch updates range from $10^{-5}|E_T|$ to $10^{-3}|E_T|$. The rank error with each approach is measured relative to ranks obtained with a reference Static PageRank run, as detailed in Section \ref{sec:measurement}.}
  \label{fig:temporal-sx-mathoverflow}
\end{figure*}

\begin{figure*}[!hbt]
  \centering
  \subfigure[Runtime on consecutive batch updates of size $10^{-5}|E_T|$]{
    \label{fig:temporal-sx-askubuntu--runtime5}
    \includegraphics[width=0.48\linewidth]{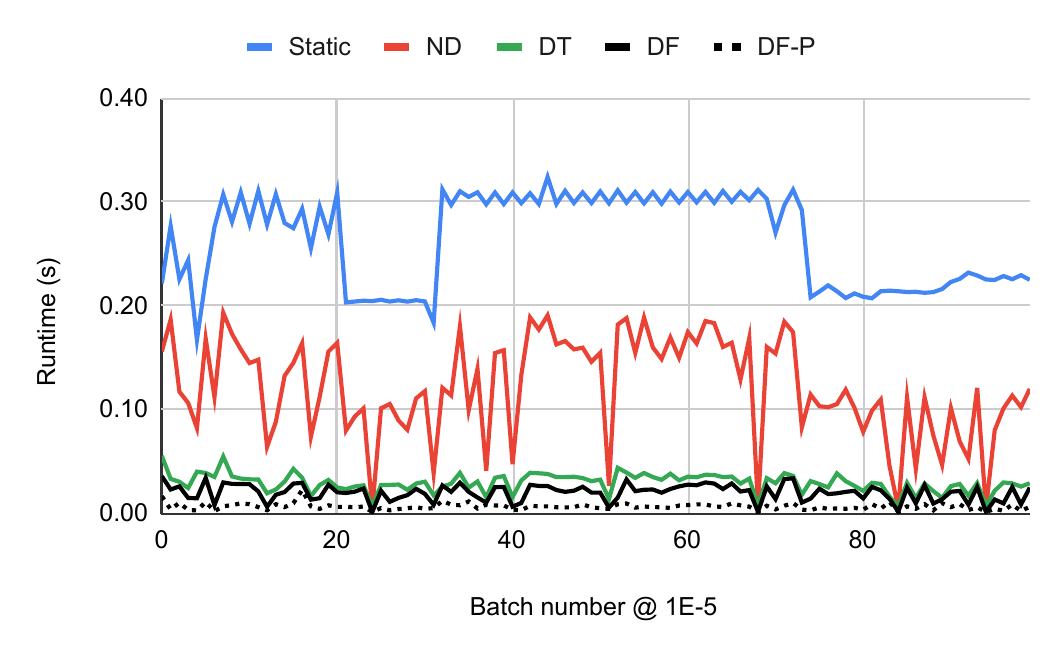}
  }
  \subfigure[Error in ranks obtained on consecutive batch updates of size $10^{-5}|E_T|$]{
    \label{fig:temporal-sx-askubuntu--error5}
    \includegraphics[width=0.48\linewidth]{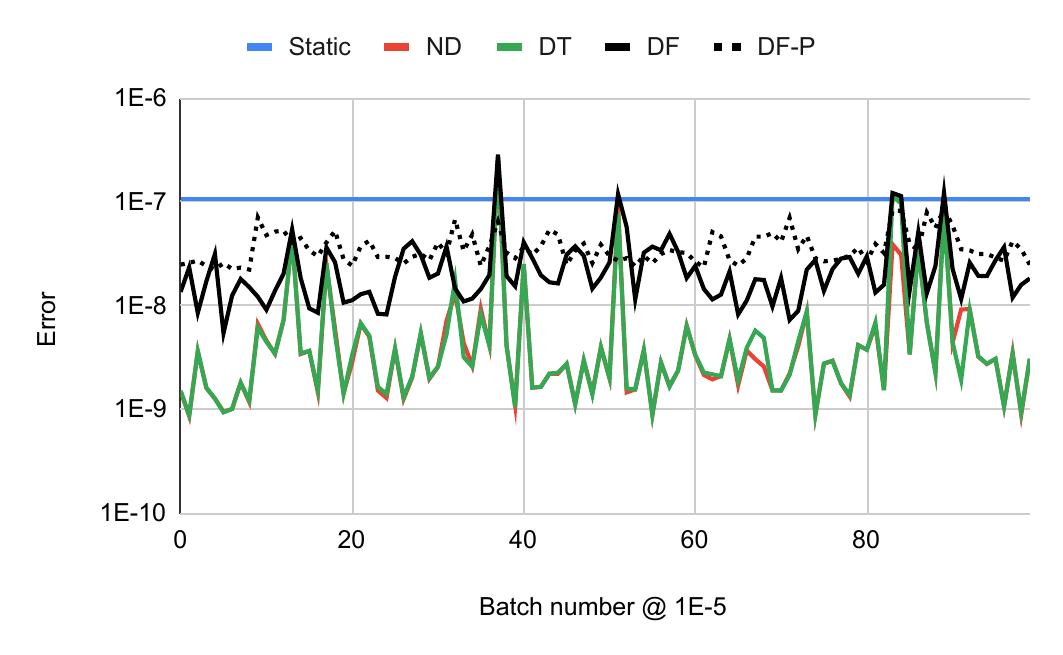}
  } \\[2ex]
  \subfigure[Runtime on consecutive batch updates of size $10^{-4}|E_T|$]{
    \label{fig:temporal-sx-askubuntu--runtime4}
    \includegraphics[width=0.48\linewidth]{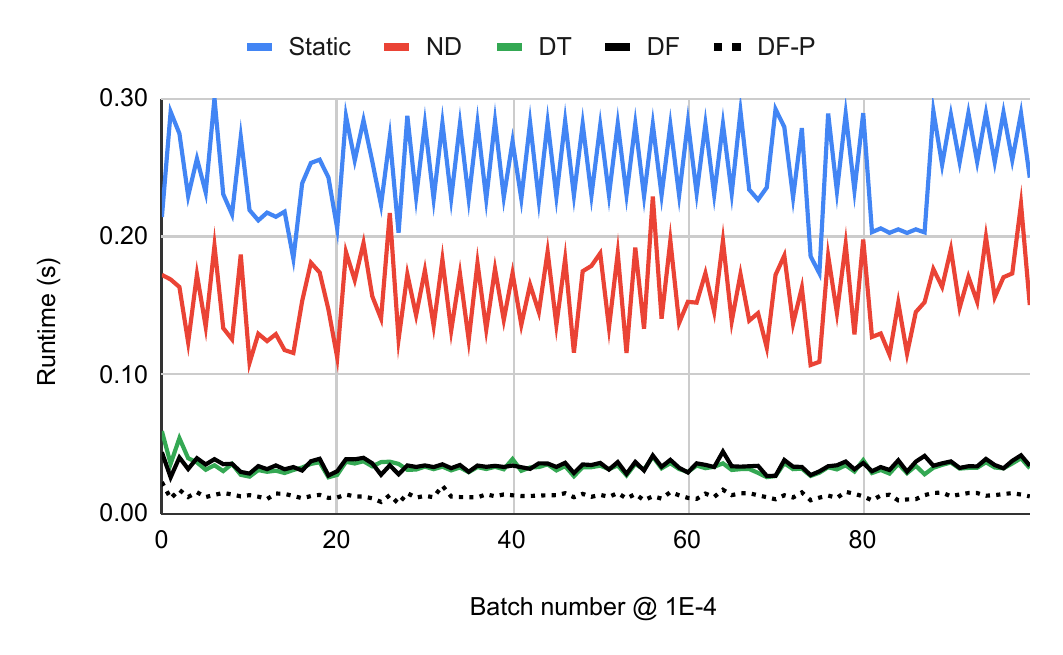}
  }
  \subfigure[Error in ranks obtained on consecutive batch updates of size $10^{-4}|E_T|$]{
    \label{fig:temporal-sx-askubuntu--error4}
    \includegraphics[width=0.48\linewidth]{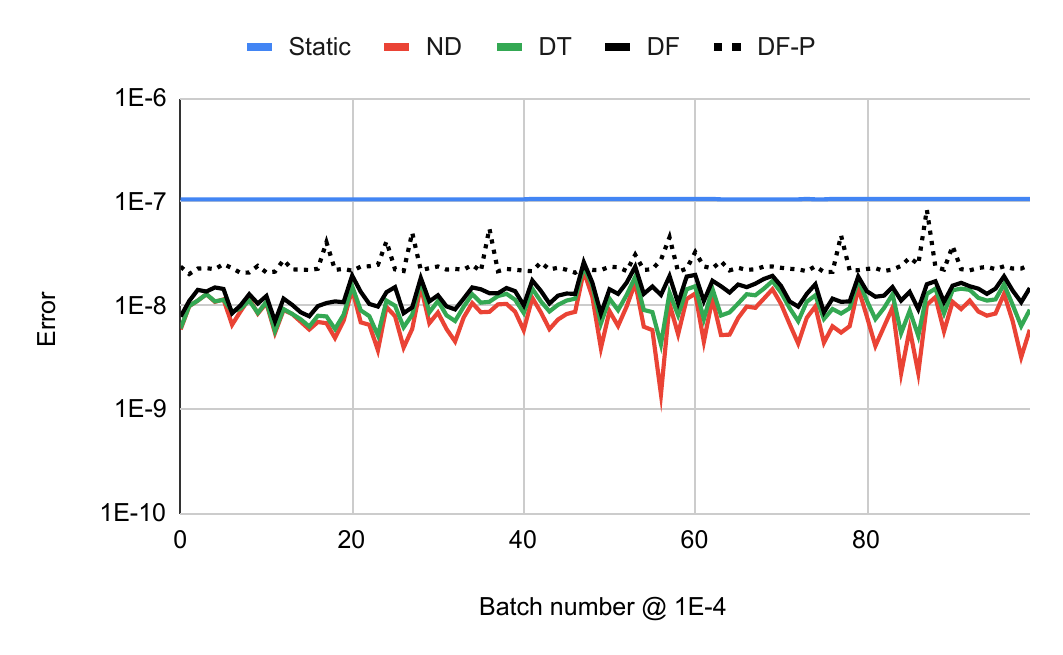}
  } \\[2ex]
  \subfigure[Runtime on consecutive batch updates of size $10^{-3}|E_T|$]{
    \label{fig:temporal-sx-askubuntu--runtime3}
    \includegraphics[width=0.48\linewidth]{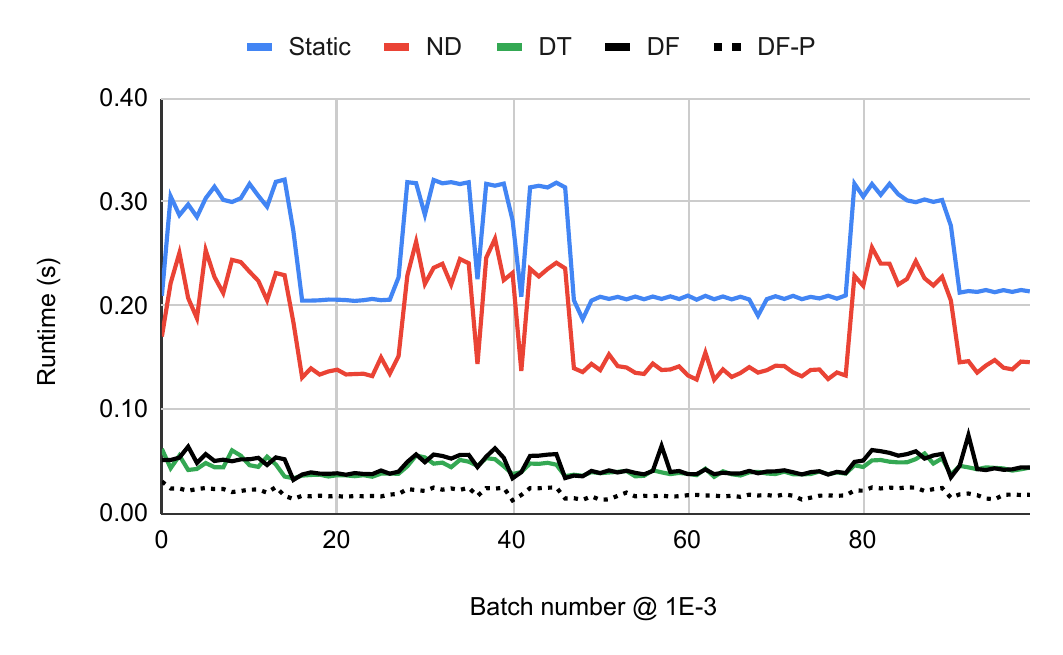}
  }
  \subfigure[Error in ranks obtained on consecutive batch updates of size $10^{-3}|E_T|$]{
    \label{fig:temporal-sx-askubuntu--error3}
    \includegraphics[width=0.48\linewidth]{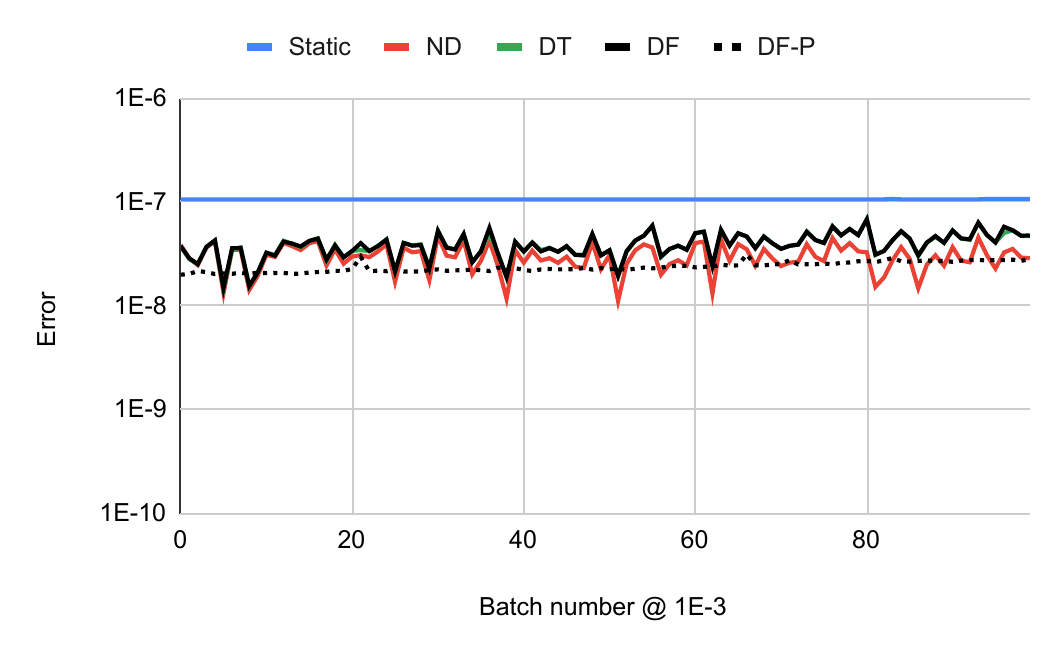}
  } \\[-2ex]
  \caption{Runtime and Error in ranks obtained with \textit{Static}, \textit{Naive-dynamic (ND)}, \textit{Dynamic Traversal (DT)}, our improved \textit{Dynamic Frontier (DF)}, and our improved \textit{Dynamic Frontier with Pruning (DF-P)} PageRank on the \textit{sx-askubuntu} dynamic graph. The size of batch updates range from $10^{-5}|E_T|$ to $10^{-3}|E_T|$. The rank error with each approach is measured relative to ranks obtained with a reference Static PageRank run, as detailed in Section \ref{sec:measurement}.}
  \label{fig:temporal-sx-askubuntu}
\end{figure*}

\begin{figure*}[!hbt]
  \centering
  \subfigure[Runtime on consecutive batch updates of size $10^{-5}|E_T|$]{
    \label{fig:temporal-sx-superuser--runtime5}
    \includegraphics[width=0.48\linewidth]{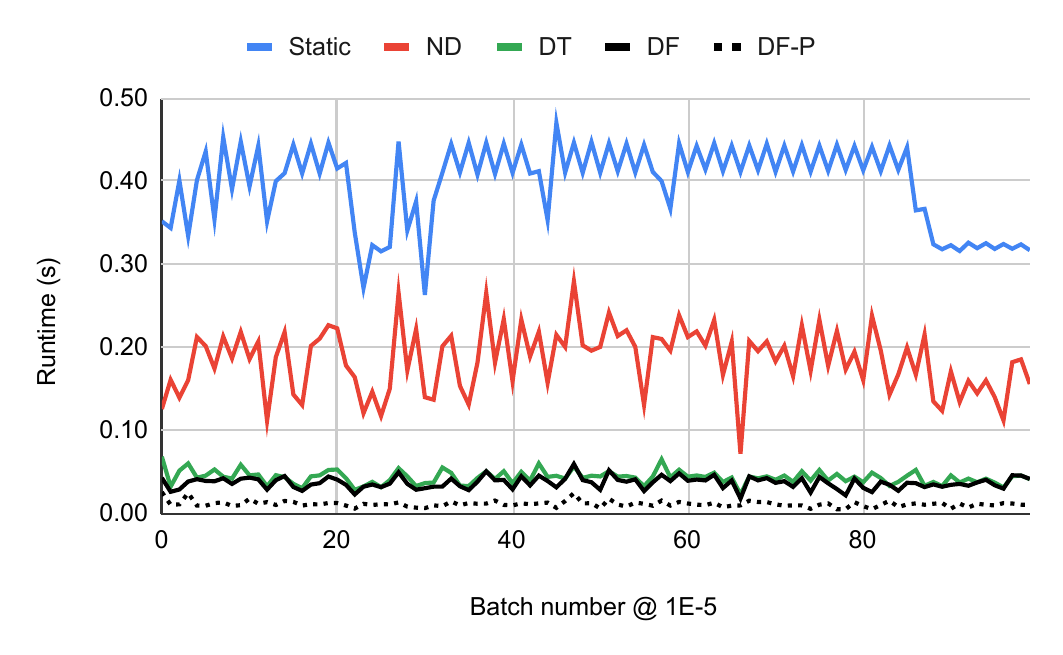}
  }
  \subfigure[Error in ranks obtained on consecutive batch updates of size $10^{-5}|E_T|$]{
    \label{fig:temporal-sx-superuser--error5}
    \includegraphics[width=0.48\linewidth]{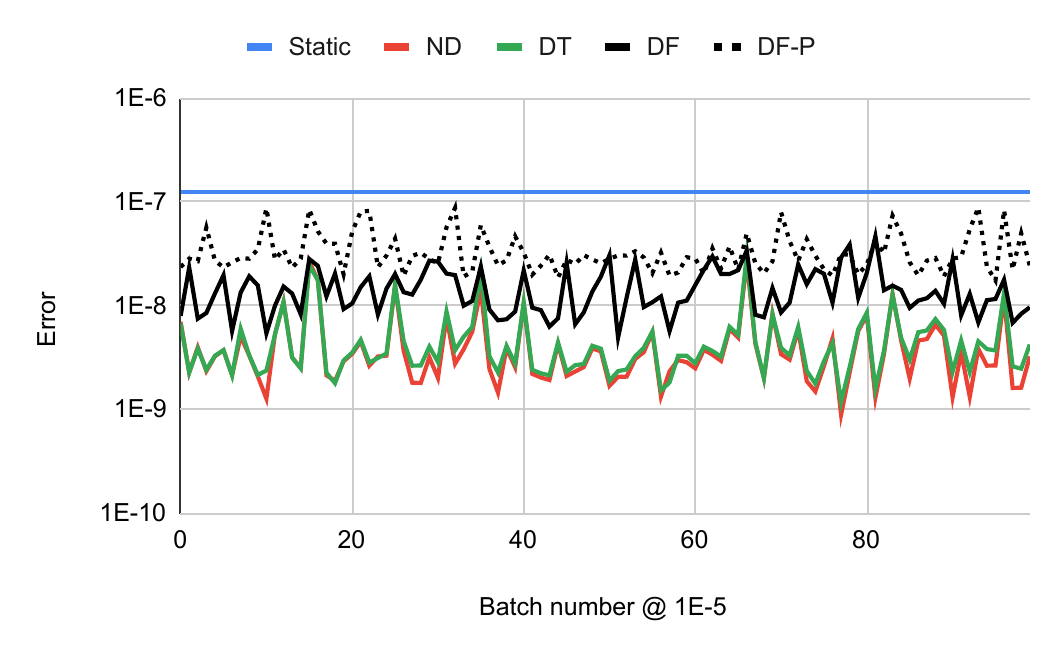}
  } \\[2ex]
  \subfigure[Runtime on consecutive batch updates of size $10^{-4}|E_T|$]{
    \label{fig:temporal-sx-superuser--runtime4}
    \includegraphics[width=0.48\linewidth]{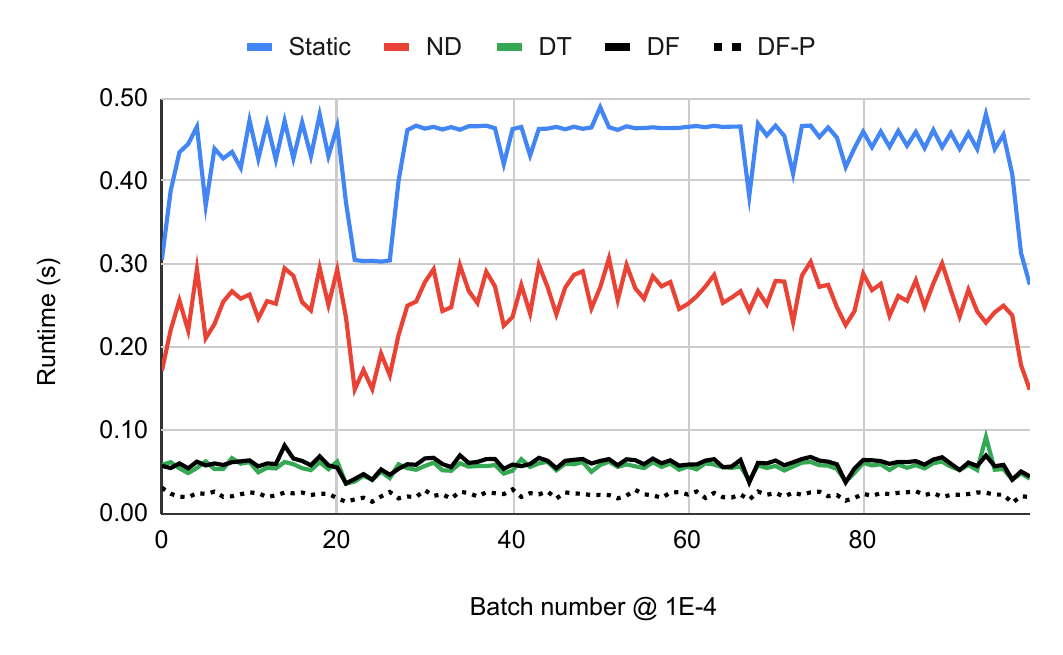}
  }
  \subfigure[Error in ranks obtained on consecutive batch updates of size $10^{-4}|E_T|$]{
    \label{fig:temporal-sx-superuser--error4}
    \includegraphics[width=0.48\linewidth]{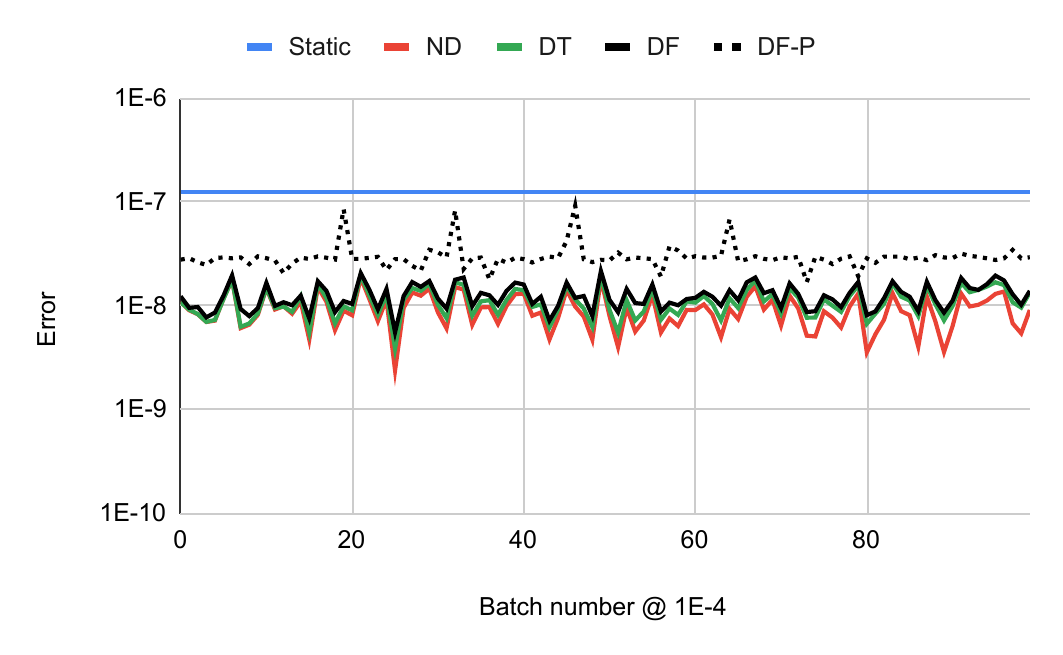}
  } \\[2ex]
  \subfigure[Runtime on consecutive batch updates of size $10^{-3}|E_T|$]{
    \label{fig:temporal-sx-superuser--runtime3}
    \includegraphics[width=0.48\linewidth]{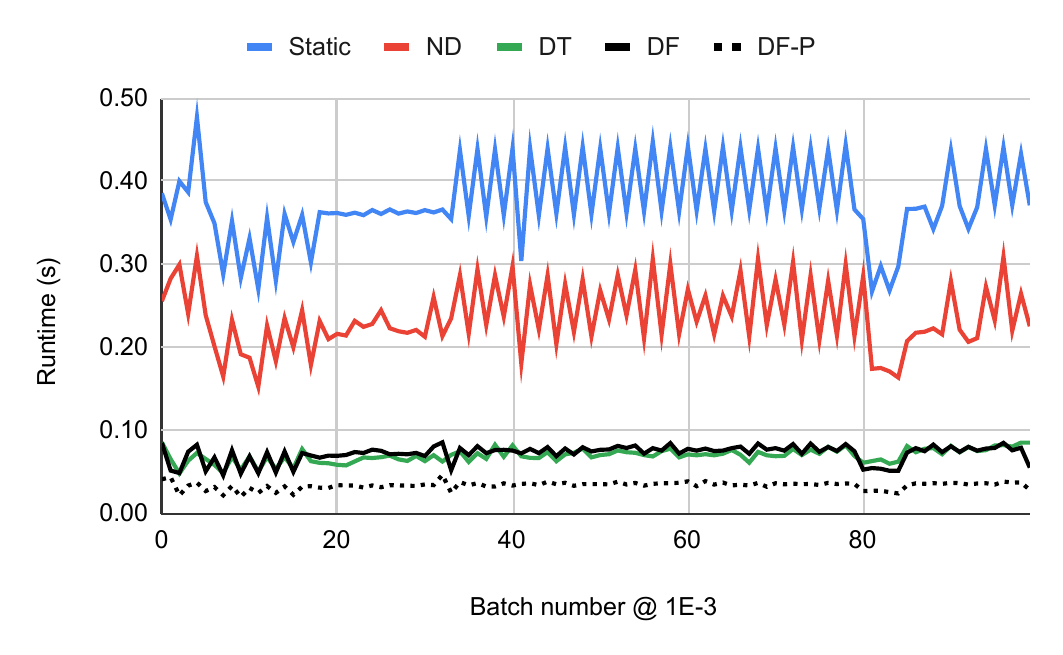}
  }
  \subfigure[Error in ranks obtained on consecutive batch updates of size $10^{-3}|E_T|$]{
    \label{fig:temporal-sx-superuser--error3}
    \includegraphics[width=0.48\linewidth]{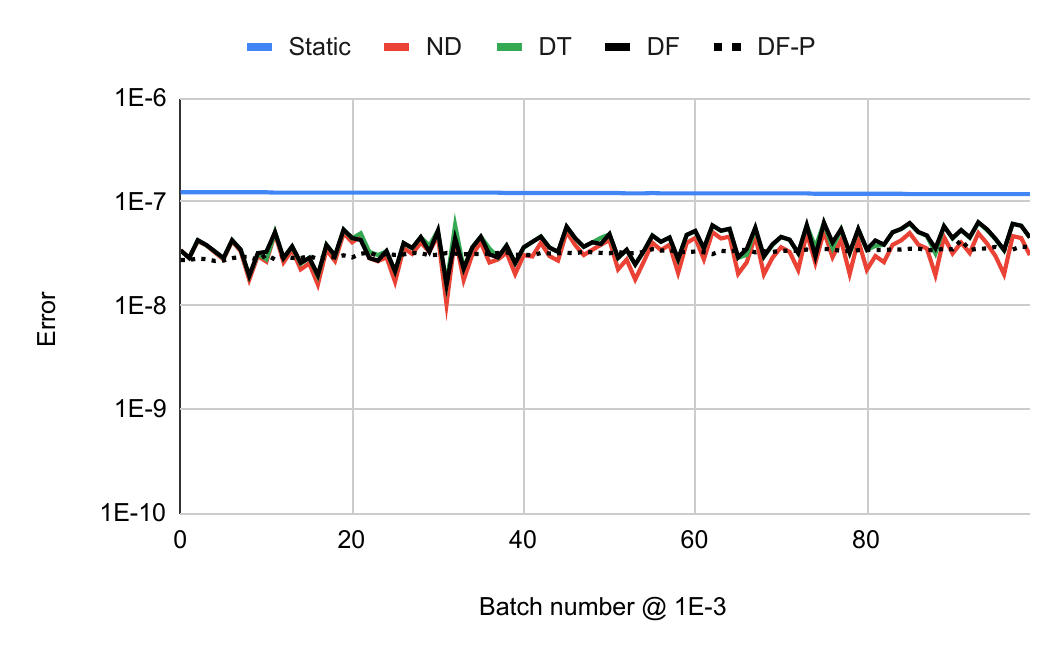}
  } \\[-2ex]
  \caption{Runtime and Error in ranks obtained with \textit{Static}, \textit{Naive-dynamic (ND)}, \textit{Dynamic Traversal (DT)}, our improved \textit{Dynamic Frontier (DF)}, and our improved \textit{Dynamic Frontier with Pruning (DF-P)} PageRank on the \textit{sx-superuser} dynamic graph. The size of batch updates range from $10^{-5}|E_T|$ to $10^{-3}|E_T|$. The rank error with each approach is measured relative to ranks obtained with a reference Static PageRank run, as detailed in Section \ref{sec:measurement}.}
  \label{fig:temporal-sx-superuser}
\end{figure*}

\begin{figure*}[!hbt]
  \centering
  \subfigure[Runtime on consecutive batch updates of size $10^{-5}|E_T|$]{
    \label{fig:temporal-wiki-talk-temporal--runtime5}
    \includegraphics[width=0.48\linewidth]{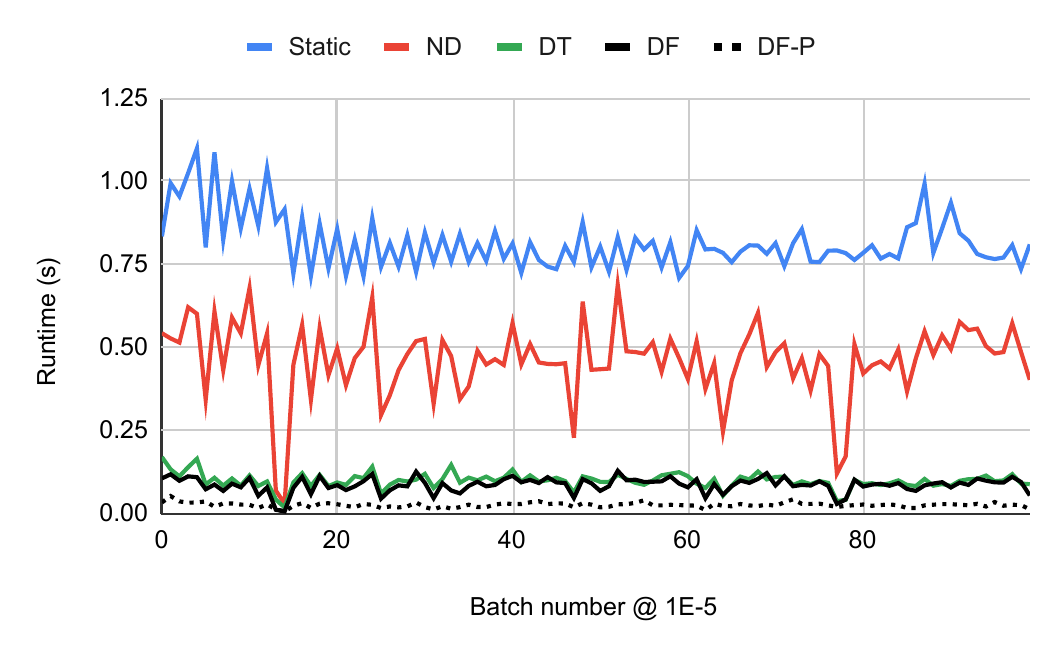}
  }
  \subfigure[Error in ranks obtained on consecutive batch updates of size $10^{-5}|E_T|$]{
    \label{fig:temporal-wiki-talk-temporal--error5}
    \includegraphics[width=0.48\linewidth]{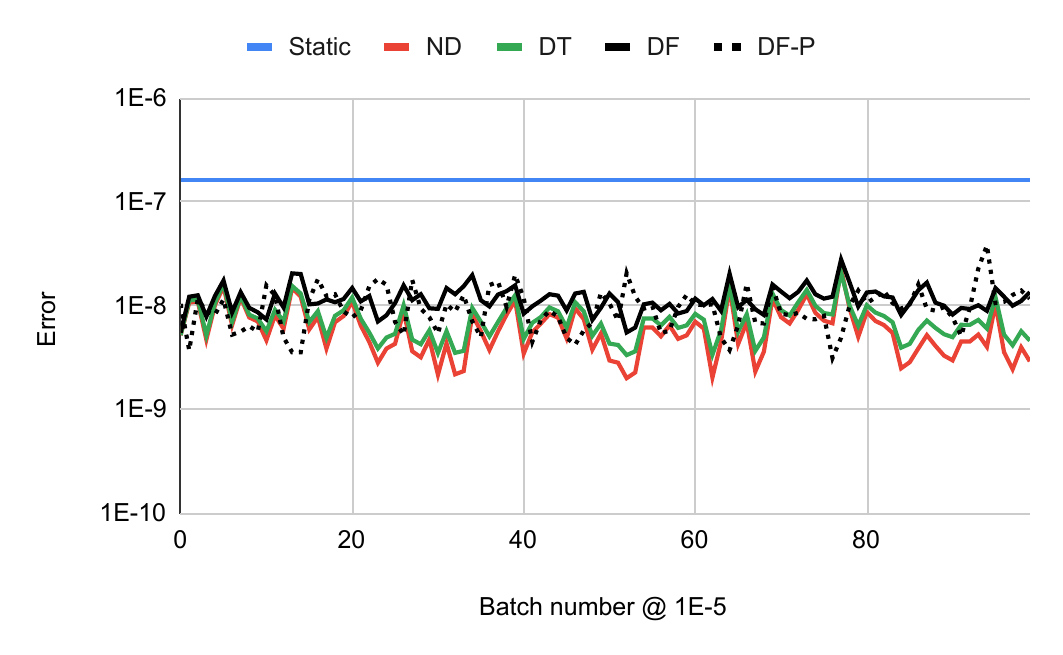}
  } \\[2ex]
  \subfigure[Runtime on consecutive batch updates of size $10^{-4}|E_T|$]{
    \label{fig:temporal-wiki-talk-temporal--runtime4}
    \includegraphics[width=0.48\linewidth]{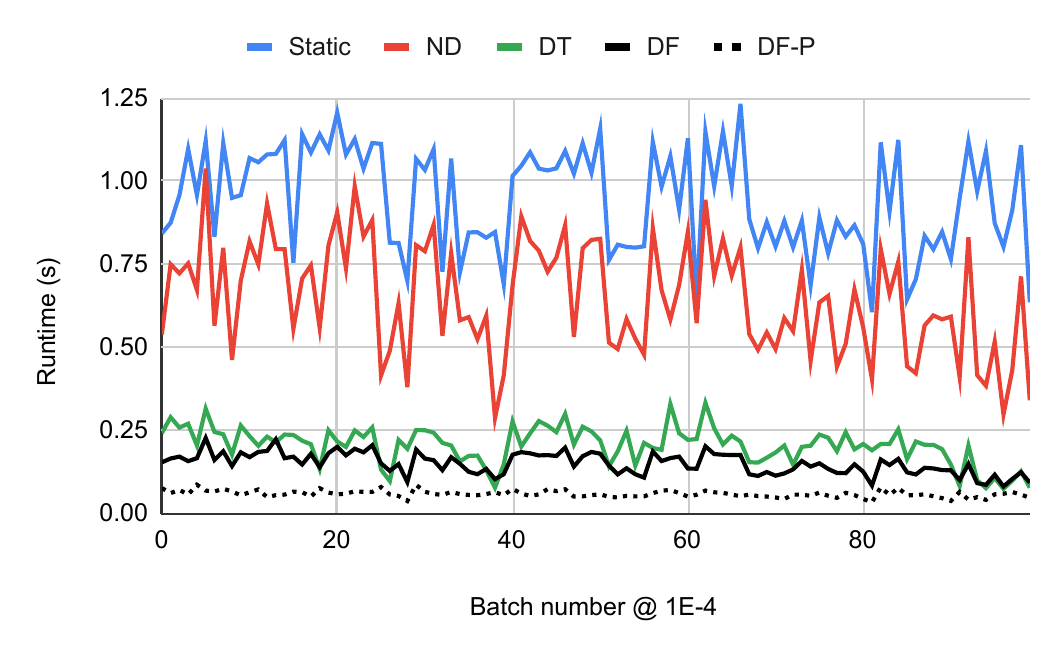}
  }
  \subfigure[Error in ranks obtained on consecutive batch updates of size $10^{-4}|E_T|$]{
    \label{fig:temporal-wiki-talk-temporal--error4}
    \includegraphics[width=0.48\linewidth]{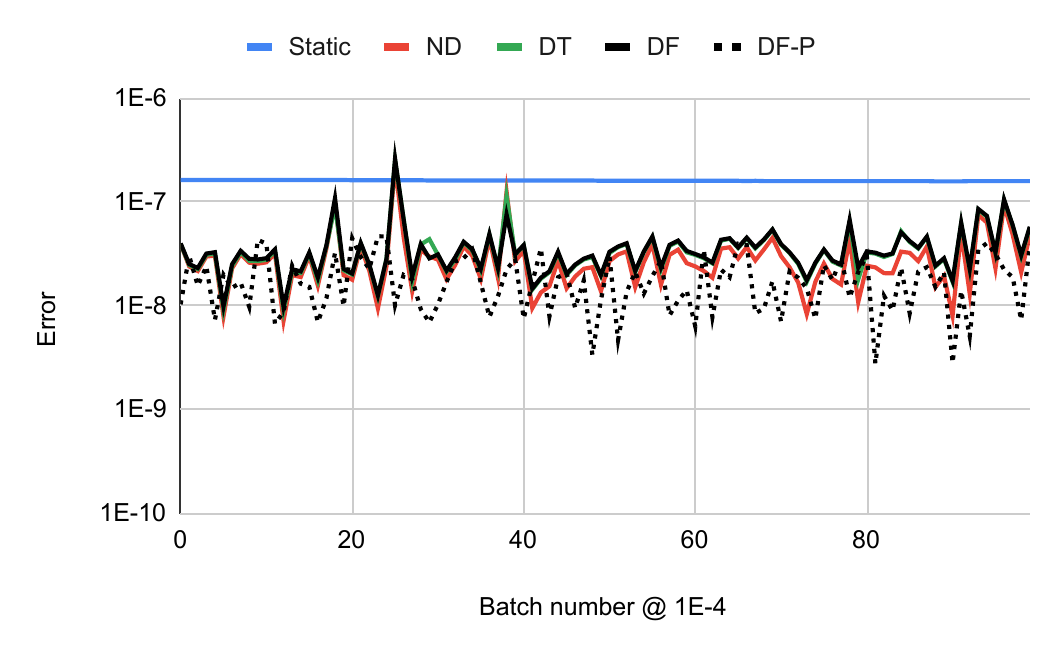}
  } \\[2ex]
  \subfigure[Runtime on consecutive batch updates of size $10^{-3}|E_T|$]{
    \label{fig:temporal-wiki-talk-temporal--runtime3}
    \includegraphics[width=0.48\linewidth]{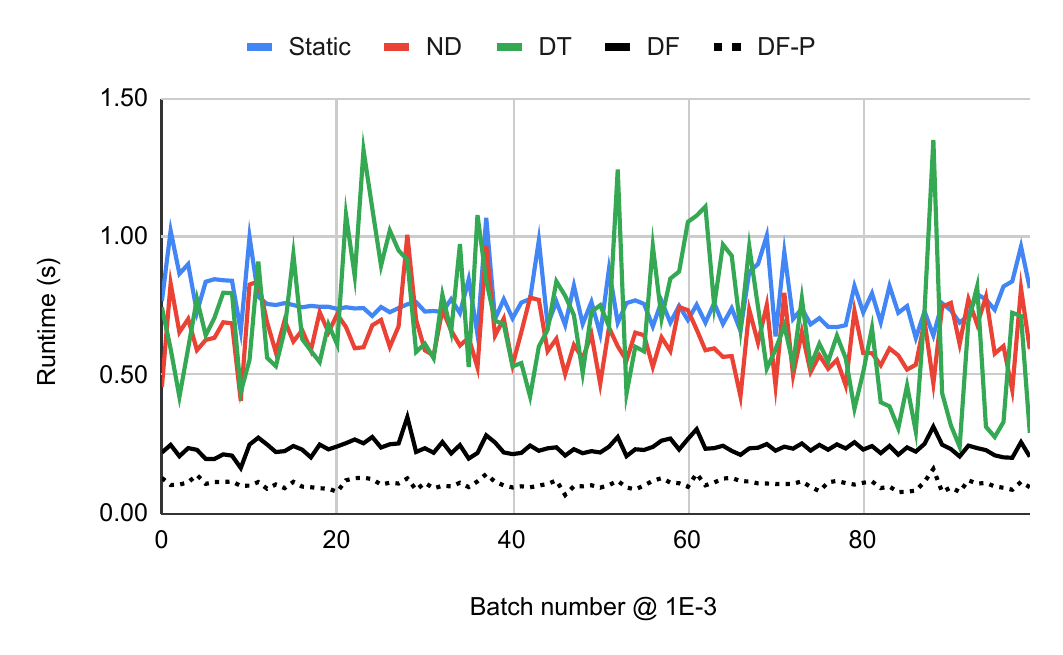}
  }
  \subfigure[Error in ranks obtained on consecutive batch updates of size $10^{-3}|E_T|$]{
    \label{fig:temporal-wiki-talk-temporal--error3}
    \includegraphics[width=0.48\linewidth]{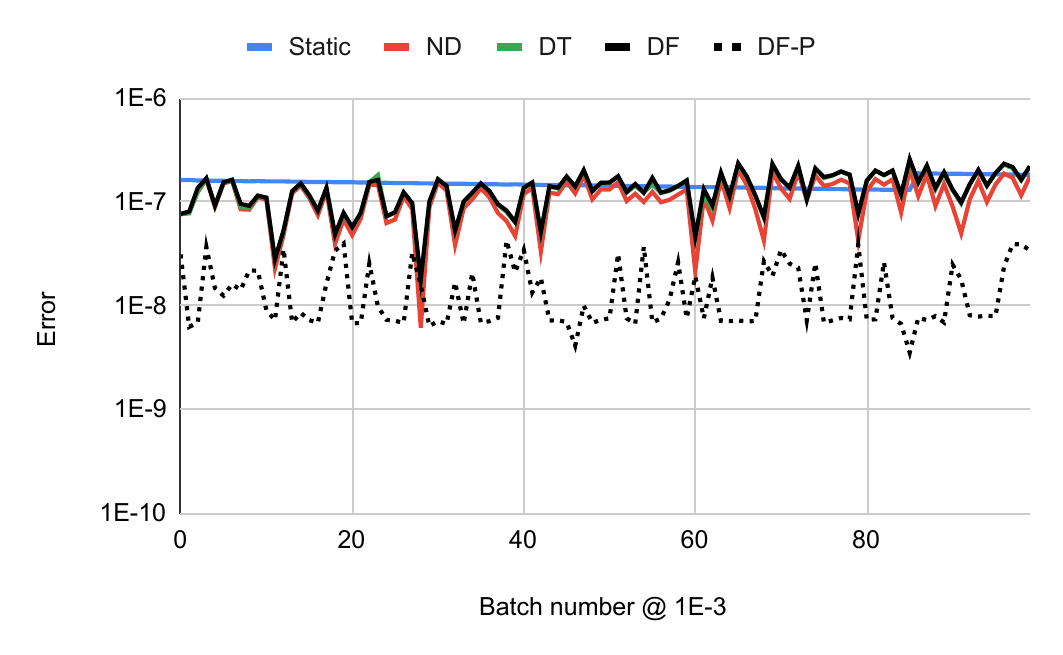}
  } \\[-2ex]
  \caption{Runtime and Error in ranks obtained with \textit{Static}, \textit{Naive-dynamic (ND)}, \textit{Dynamic Traversal (DT)}, our improved \textit{Dynamic Frontier (DF)}, and our improved \textit{Dynamic Frontier with Pruning (DF-P)} PageRank on the \textit{wiki-talk-temporal} dynamic graph. The size of batch updates range from $10^{-5}|E_T|$ to $10^{-3}|E_T|$. The rank error with each approach is measured relative to ranks obtained with a reference Static PageRank run, as detailed in Section \ref{sec:measurement}.}
  \label{fig:temporal-wiki-talk-temporal}
\end{figure*}

\begin{figure*}[!hbt]
  \centering
  \subfigure[Runtime on consecutive batch updates of size $10^{-5}|E_T|$]{
    \label{fig:temporal-sx-stackoverflow--runtime5}
    \includegraphics[width=0.48\linewidth]{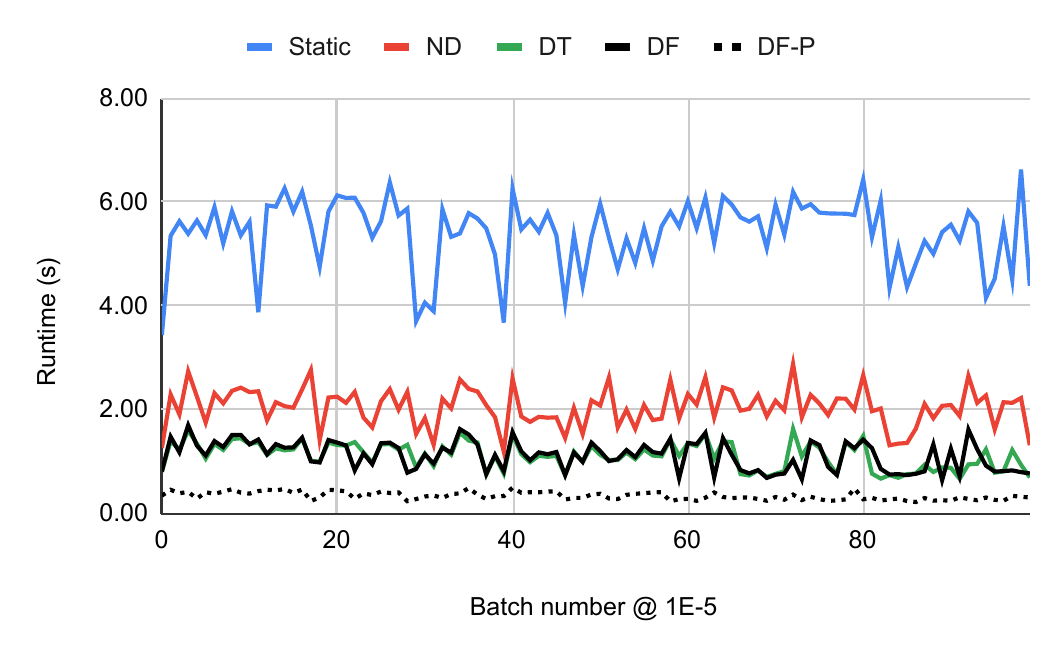}
  }
  \subfigure[Error in ranks obtained on consecutive batch updates of size $10^{-5}|E_T|$]{
    \label{fig:temporal-sx-stackoverflow--error5}
    \includegraphics[width=0.48\linewidth]{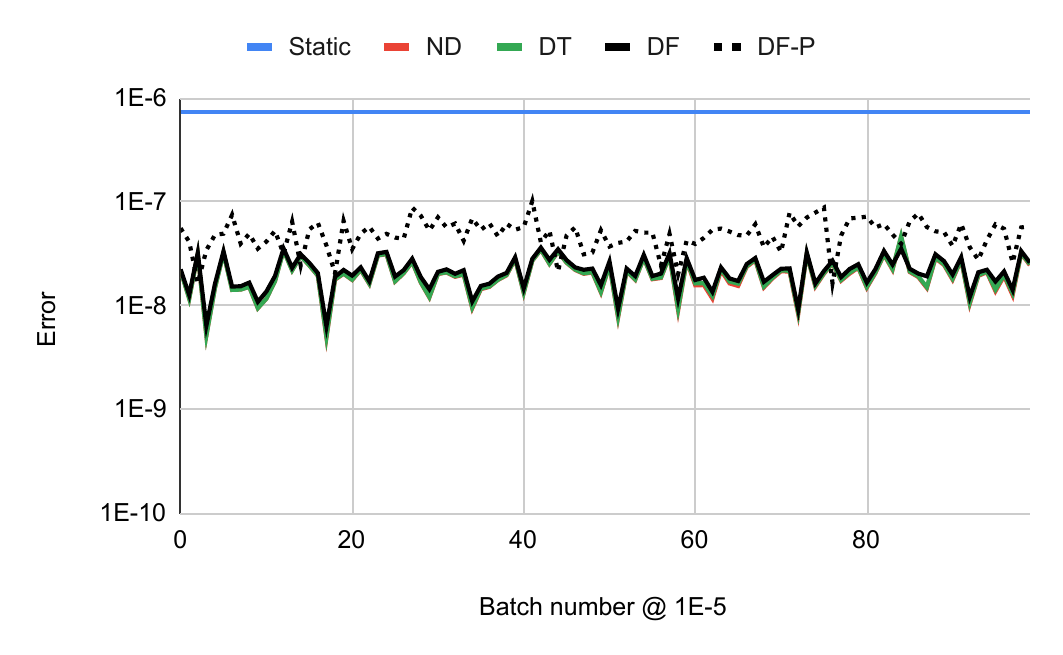}
  } \\[2ex]
  \subfigure[Runtime on consecutive batch updates of size $10^{-4}|E_T|$]{
    \label{fig:temporal-sx-stackoverflow--runtime4}
    \includegraphics[width=0.48\linewidth]{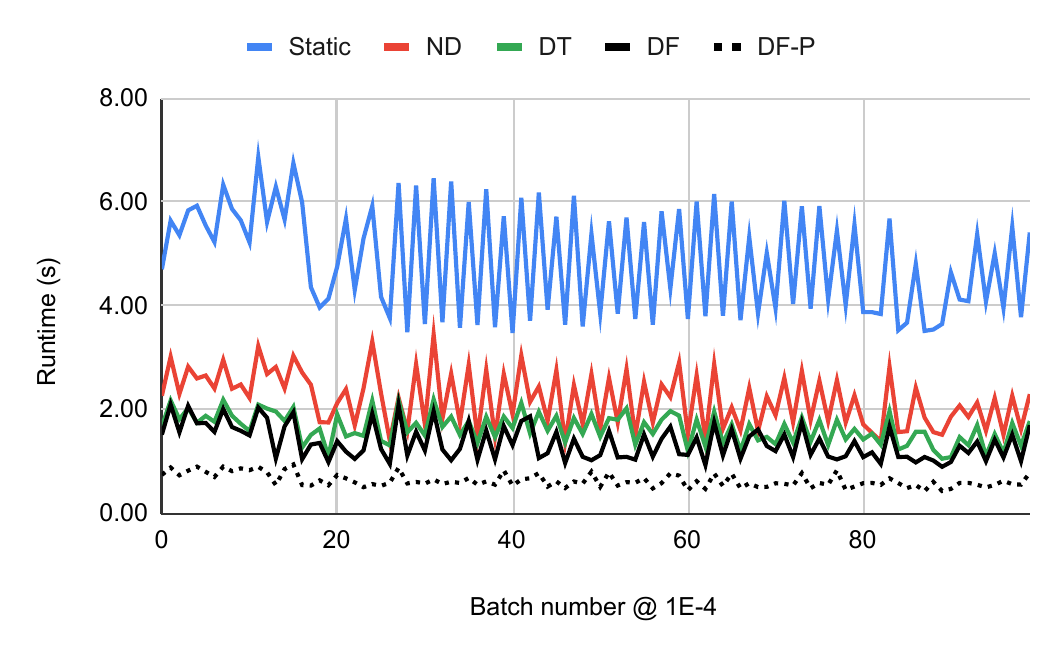}
  }
  \subfigure[Error in ranks obtained on consecutive batch updates of size $10^{-4}|E_T|$]{
    \label{fig:temporal-sx-stackoverflow--error4}
    \includegraphics[width=0.48\linewidth]{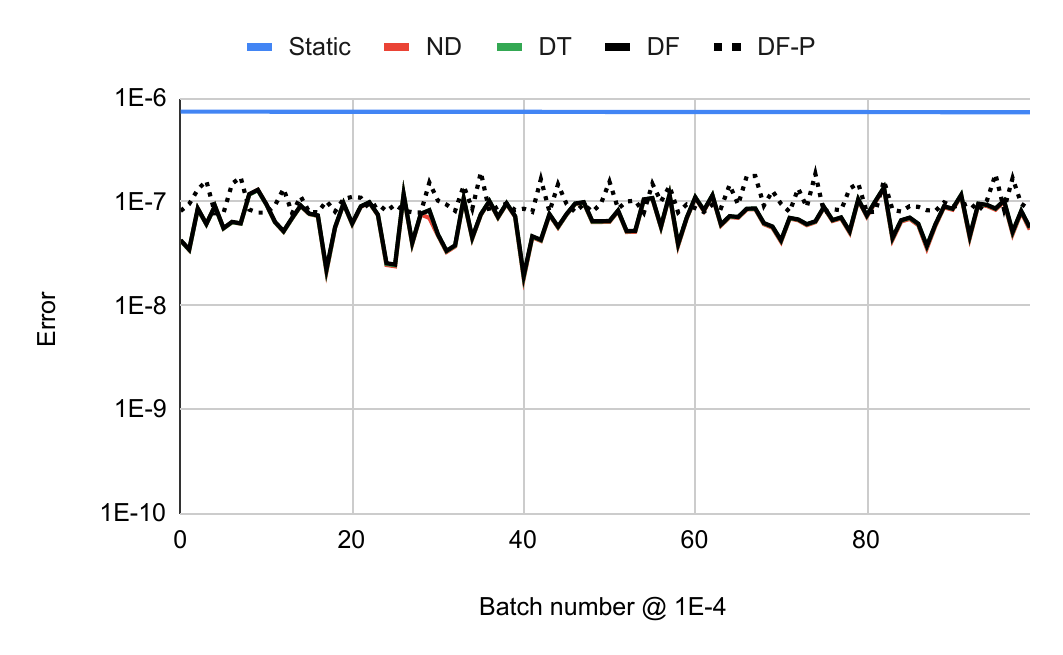}
  } \\[2ex]
  \subfigure[Runtime on consecutive batch updates of size $10^{-3}|E_T|$]{
    \label{fig:temporal-sx-stackoverflow--runtime3}
    \includegraphics[width=0.48\linewidth]{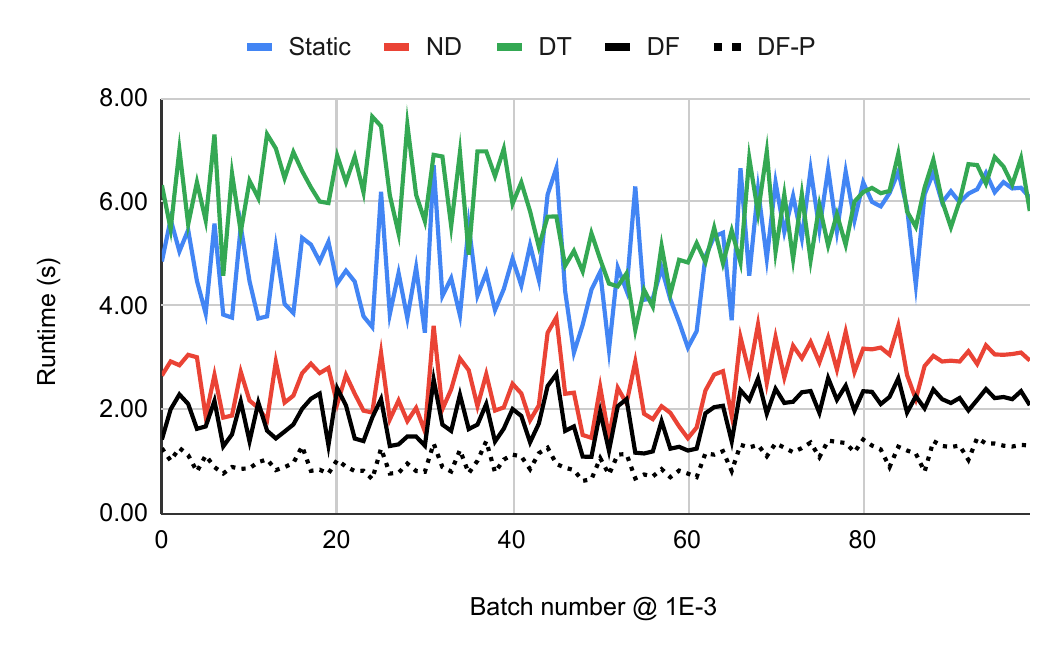}
  }
  \subfigure[Error in ranks obtained on consecutive batch updates of size $10^{-3}|E_T|$]{
    \label{fig:temporal-sx-stackoverflow--error3}
    \includegraphics[width=0.48\linewidth]{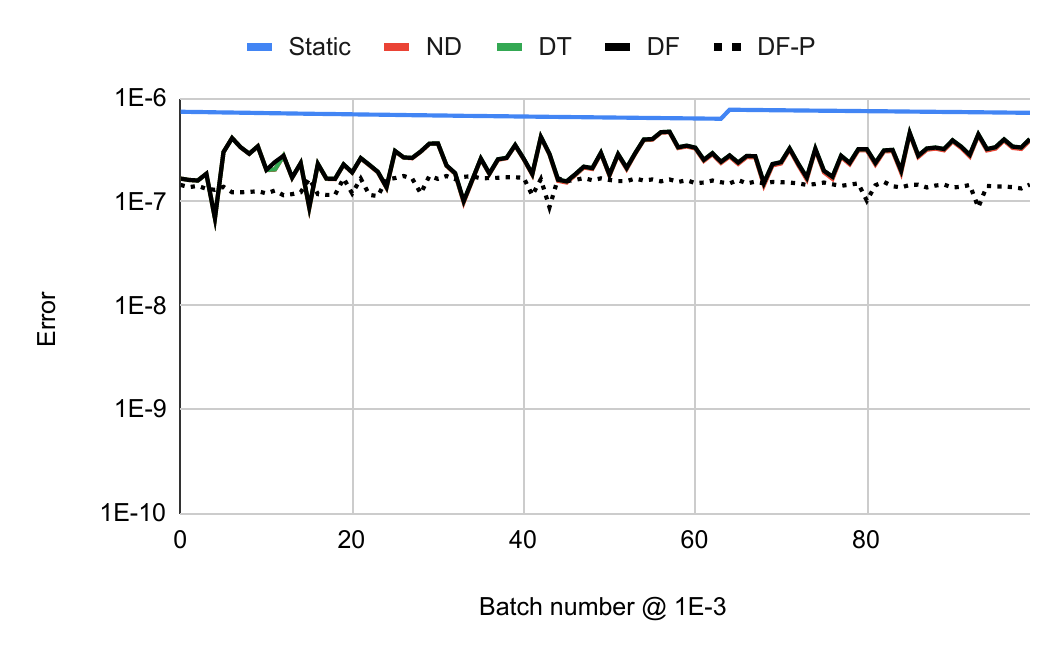}
  } \\[-2ex]
  \caption{Runtime and Error in ranks obtained with \textit{Static}, \textit{Naive-dynamic (ND)}, \textit{Dynamic Traversal (DT)}, our improved \textit{Dynamic Frontier (DF)}, and our improved \textit{Dynamic Frontier with Pruning (DF-P)} PageRank on the \textit{sx-stackoverflow} dynamic graph. The size of batch updates range from $10^{-5}|E_T|$ to $10^{-3}|E_T|$. The rank error with each approach is measured relative to ranks obtained with a reference Static PageRank run, as detailed in Section \ref{sec:measurement}.}
  \label{fig:temporal-sx-stackoverflow}
\end{figure*}

\begin{figure*}[hbtp]
  \centering
  \subfigure[Overall result]{
    \label{fig:8020-runtime--mean}
    \includegraphics[width=0.38\linewidth]{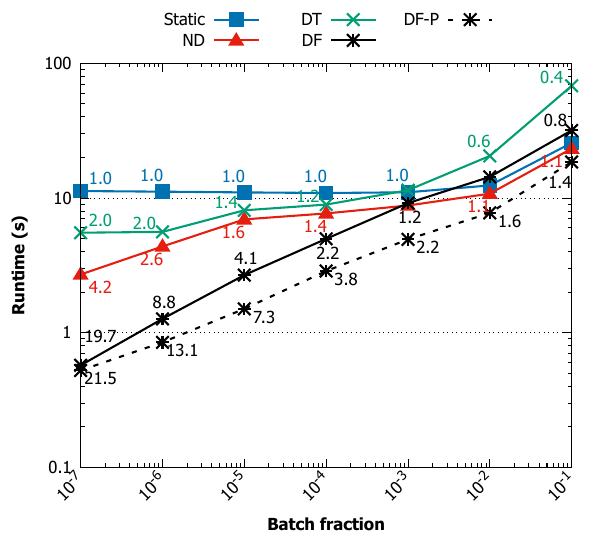}
  }
  \subfigure[Results on each graph]{
    \label{fig:8020-runtime--all}
    \includegraphics[width=0.58\linewidth]{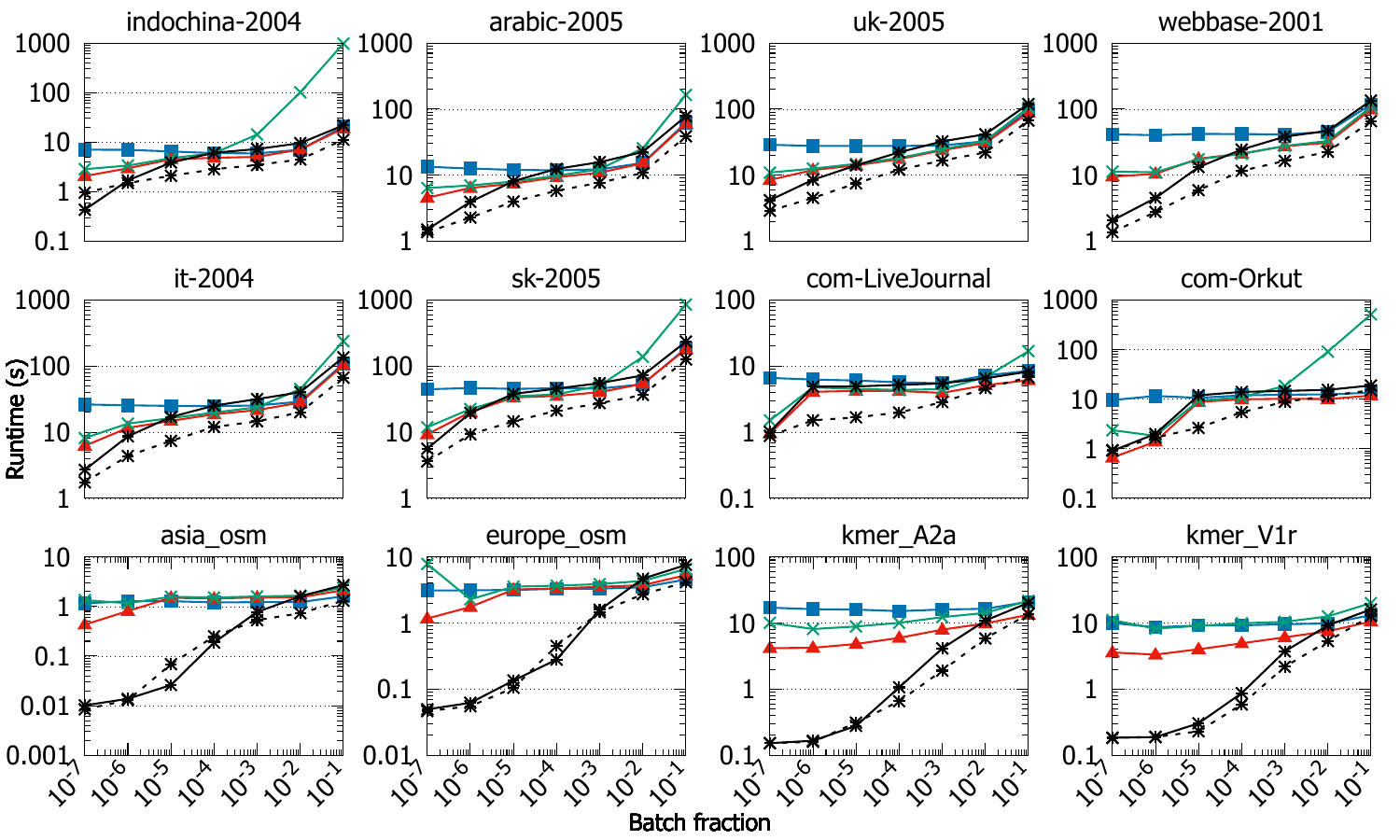}
  } \\[-1ex]
  \caption{Runtime (logarithmic scale) of \textit{Static}, \textit{Naive-dynamic (ND)}, \textit{Dynamic Traversal (DT)}, our improved \textit{Dynamic Frontier (DF)}, and \textit{Dynamic Frontier with Pruning (DF-P)} PageRank on large (static) graphs with generated random batch updates, on batch updates of size $10^{-7}|E|$ to $0.1|E|$ in multiples of $10$. The updates include $80\%$ edge insertions and $20\%$ edge deletions, simulating realistic changes upon a dynamic graph. The subfigure on the right illustrates the runtime of each approach for each graph in the dataset, while the subfigure of the left presents overall runtimes (using geometric mean for consistent scaling across graphs). In addition, the speedup of each approach, relative to Static PageRank, is labeled on respective lines.}
  \label{fig:8020-runtime}
\end{figure*}

\begin{figure*}[hbtp]
  \centering
  \subfigure[Overall result]{
    \label{fig:8020-error--mean}
    \includegraphics[width=0.38\linewidth]{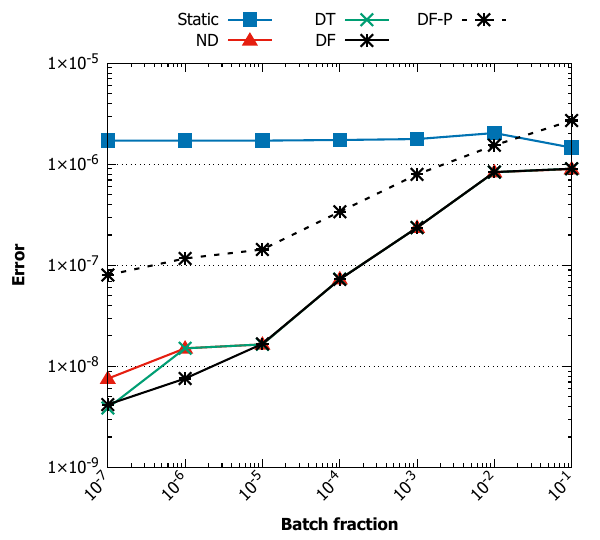}
  }
  \subfigure[Results on each graph]{
    \label{fig:8020-error--all}
    \includegraphics[width=0.58\linewidth]{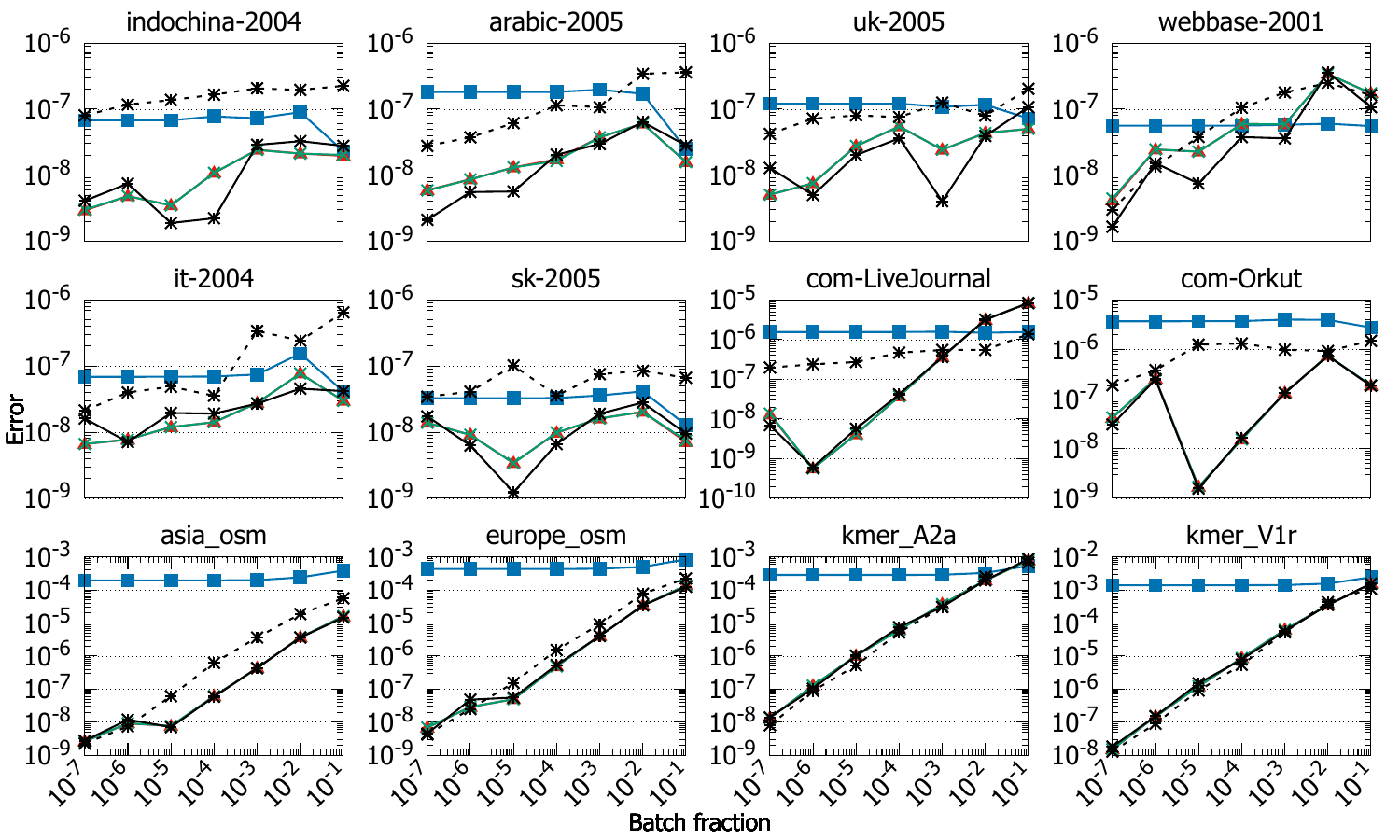}
  } \\[-1ex]
  \caption{Error comparison of \textit{Static}, \textit{Naive-dynamic (ND)}, \textit{Dynamic Traversal (DT)}, our improved \textit{Dynamic Frontier (DF)}, and \textit{Dynamic Frontier with Pruning (DF-P)} PageRank on large (static) graphs with generated random batch updates, relative to a Reference Static PageRank (see Section \ref{sec:measurement}), using $L1$-norm. The size of batch updates range from $10^{-7} |E|$ to $0.1 |E|$ in multiples of $10$ (logarithmic scale), consisting of $80\%$ edge insertions and $20\%$ edge deletions to simulate realistic dynamic graph updates. The right subfigure depicts the error for each approach in relation to each graph, while the left subfigure showcases overall errors using geometric mean for consistent scaling across graphs.}
  \label{fig:8020-error}
\end{figure*}

\clearpage

\section{Appendix}

\subsection{Derivation of Closed loop formula for Rank calculation towards Dynamic Frontier with Pruning (DF-P) PageRank}
\label{sec:pr-prune-derivation}

We proceed to derive the closed-loop formula for rank calculation with DF-P PageRank. As outlined in Sections \ref{sec:dataset} and \ref{sec:batch-generation}, self-loops are added to each vertex to circumvent the need for a global teleport rank computation in every iteration, thus reducing overhead. In DF-P PageRank, our aim is to skip the computation of ranks for vertices likely to have already converged. However, the existence of self-loops causes a delay in vertex rank convergence due to the immediate recursive nature they introduce. For instance, if the ranks of all in-neighbors of a vertex have already converged, the presence of self-loops inhibits the convergence of the vertex's rank in a single iteration. Nevertheless, we can mitigate this convergence issue by employing a closed-loop formula for the rank calculation of each vertex.

To achieve this, let us denote $r_0$ as the initial rank of a vertex $v$, $\alpha$ as the damping factor, $c = \sum_{u \in G.in(v)\ |\ u \neq v} \frac{R[u]}{|G.out(u)|}$ as the total rank contribution from its in-neighbors (excluding itself), $d = |G.out(v)|$ as its out-degree, and $C_0$ as $1 - \alpha/|V|$. Given the assumption that the rank contribution of its in-neighbors remains constant, the rank of $v$ after one iteration can be expressed as:

\begin{flalign*}
  r_1 & = \alpha (c + \frac{r_0}{d}) + C_0 && \\
      & = \alpha c + \alpha \frac{r_0}{d} + C_0 && \\
\end{flalign*}

\noindent
After the second iteration, the rank of the vertex would be:

\begin{flalign*}
  r_2 & = \alpha (c + \frac{r_1}{d}) + C_0 && \\
      & = \alpha (c + \frac{1}{d} (\alpha c + \alpha \frac{r_0}{d} + C_0)) + C_0 && \\
      & = \alpha c + \alpha^2 \frac{c}{d} + \alpha^2 \frac{r_0}{d^2} + \alpha \frac{C_0}{d} + C_0 &&
\end{flalign*}

\noindent
Following the third iteration, the vertex's rank would be:

\begin{flalign*}
  r_3 & = \alpha (c + \frac{r_2}{d}) + C_0 && \\
      & = \alpha (c + \frac{1}{d} (\alpha c + \alpha^2 \frac{c}{d} + \alpha^2 \frac{r_0}{d^2} + \alpha \frac{C_0}{d} + C_0) + C_0 && \\
      & = \alpha c + \alpha^2 \frac{c}{d} + \alpha^3 \frac{c}{d^2} + \alpha^3 \frac{r_0}{d^3} + \alpha^2 \frac{C_0}{d^2} + \alpha \frac{C_0}{d} + C_0 && \\
\end{flalign*}

\noindent
Expanding this to an infinite number of iterations, the vertex's final rank would be:

\begin{flalign*}
  r_\infty & = \frac{\alpha c}{1 - \alpha / d} + \frac{C_0}{1 - \alpha / d} && \\
           & = \frac{1}{1 - \alpha / d} (\alpha c + C_0)
\end{flalign*}

\noindent
Hence, the closed-loop formula for calculating the rank of a vertex $v$ in DF-P PageRank is:

\begin{flalign}
  R[v] & = \frac{1}{1 - \alpha / |G.out(v)|} \left(\alpha K + \frac{1 - \alpha}{|V|}\right) && \\
    \text{where, } K & = \left(\sum_{u \in G.in(v)} \frac{R[u]}{|G.out(u)|}\right) - \frac{R[v]}{|G.out(v)|}
\end{flalign}

\begin{table}[hbtp]
  \centering
  \caption{List of $12$ graphs sourced from the SuiteSparse Matrix Collection \cite{suite19}, where directed graphs are indicated with $*$. Here, $|V|$ denotes the number of vertices, $|E|$ represents the number of edges (inclusive of self-loops), and $D_{avg}$ represents the average degree.}
  \label{tab:dataset-large}
  \begin{tabular}{|c||c|c|c|c|}
    \toprule
    \textbf{Graph} &
    \textbf{\textbf{$|V|$}} &
    \textbf{\textbf{$|E|$}} &
    \textbf{\textbf{$D_{avg}$}} \\
    \midrule
    \multicolumn{4}{|c|}{\textbf{Web Graphs (LAW)}} \\ \hline
    indochina-2004$^*$ & 7.41M & 199M & 26.8 \\ \hline  
    arabic-2005$^*$ & 22.7M & 654M & 28.8 \\ \hline  
    uk-2005$^*$ & 39.5M & 961M & 24.3 \\ \hline  
    webbase-2001$^*$ & 118M & 1.11B & 9.4 \\ \hline  
    it-2004$^*$ & 41.3M & 1.18B & 28.5 \\ \hline  
    sk-2005$^*$ & 50.6M & 1.98B & 39.1 \\ \hline  
    \multicolumn{4}{|c|}{\textbf{Social Networks (SNAP)}} \\ \hline
    com-LiveJournal & 4.00M & 73.4M & 18.3 \\ \hline  
    com-Orkut & 3.07M & 237M & 77.3 \\ \hline  
    \multicolumn{4}{|c|}{\textbf{Road Networks (DIMACS10)}} \\ \hline
    asia\_osm & 12.0M & 37.4M & 3.1 \\ \hline  
    europe\_osm & 50.9M & 159M & 3.1 \\ \hline  
    \multicolumn{4}{|c|}{\textbf{Protein k-mer Graphs (GenBank)}} \\ \hline
    kmer\_A2a & 171M & 531M & 3.1 \\ \hline  
    kmer\_V1r & 214M & 679M & 3.2 \\ \hline  
  \bottomrule
  \end{tabular}
\end{table}

\end{document}